\documentclass[11pt,authoryear]{article}

\usepackage{epsf}
\usepackage{natbib}

\newcommand{\be}{\begin{equation}}
\newcommand{\ee}{\end{equation}}

\begin{document}


\setcounter{topnumber}{2}
\setcounter{bottomnumber}{2}
\setcounter{totalnumber}{4}
\renewcommand{\topfraction}{0.9}
\renewcommand{\bottomfraction}{0.9}
\renewcommand{\textfraction}{0.1}
\renewcommand{\floatpagefraction}{0.8}

\title{A note on Keen's model: The limits of Schumpeter's ``Creative Destruction''}

\author{Glenn Ierley \\
U C San Diego, Professor Emeritus\\
grierley@ucsd.edu}

\maketitle

\begin{abstract}
This paper presents a general solution for a recent model by Keen for 
endogenous money creation. The solution provides an analytic framework
that explains all significant dynamical features of Keen's model and 
their parametric dependence, including an exact result for both the 
period and subsidence rate of the Great Moderation. It emerges that 
Keen's model has just two possible long term solutions: stable
growth or terminal collapse. While collapse can come about immediately
from economies that are nonviable by virtue of unsuitable parameters or
initial conditions, in general the collapse is preceded by an interval of
exponential growth. In first approximation, the duration of that
exponential growth is half a period of a sinusoidal oscillation.  The
period is determined by reciprocal of the imaginary part of one root of a
certain quintic polynomial. The real part of the same root determines the
rate of growth of the economy. The coefficients of that polynomial depend
in a complicated way upon the numerous parameters in the problem and so,
therefore, the pattern of roots. For a favorable choice of parameters, the
salient root is purely real. This is the circumstance that admits the second
possible long term solution, that of indefinite stable growth, i.e.\ an infinite
period.
\end{abstract}

\bigskip
\noindent
{\bf Keywords:}
Keen model, Financial instability hypothesis, Asymptotic solution, Nonlinear dynamics, Endogenous money
\bigskip

\noindent
{\bf JEL:} B50, C62, C63, E12, E47

\section{Introduction}
\label{sect:intro}

In \cite{Keen:jpke}, \cite{Keen:book}, \cite{Keen:eap}, 
\cite{Keen:jebo}, Steve Keen has introduced a
mathematical formalism that addresses the Financial Instability
Hypothesis of \cite{minsky} in the form of  low order nonlinear systems of
differential equations.  When parameters are set to values consistent with
estimates from macroeconomic modeling, models of this kind developed by Keen can
exhibit instability reminiscent of that studied in chaos theory and, more
important, instability much like that seen in the Great Moderation
followed by the rapid economic spiral downward commencing in 2008.

Nonlinear systems often have a rich spectrum of behaviors as the governing
parameters are changed and here, where there are ten or more parameters,
exploring all possible such behaviors would appear a daunting task,
although perhaps not hopelessly so in that models relevant to the real
world occupy only a modest range of the otherwise mathematically unbounded
parameter space. One way to classify such systems is to ask for a given
set of parameters what are the possible long time characteristic
(``asymptotic") behaviors of the system for a wide range of initial
conditions (though here, too, considering only ``realistic", rather than
all possible, initial conditions).

To speak now of the specific model explored here, one such behavior is
that noted; collapse in the long term, with various economic quantities
decaying exponentially with time.  It is significant that this collapse is
always a possible terminal solution, regardless of parameter
settings.\footnote{In this respect the model considered here is incomplete
  because recovery does eventually take place; growth resumes.}
Nonetheless, this collapse is {\sl not} always the invariable long time
characteristic behavior for all initial conditions of interest. Rather, for only
certain parameter regimes does one realize this limit irrespective of
initial conditions. For other parameter regimes, this collapse can only
come about for an exceedingly small range of possible starting conditions
and quite another long term solution is the tendency approached under most
economic conditions: a stable, steadily growing, ``Schumpeter economy'' as
Keen has termed this model.

Though Keen's model is an intricate ninth order nonlinear system, quite
remarkably it admits a {\em general} asymptotic solution --- that is, one
with nine free parameters --- whose form depends upon particular roots
that emerge from solving an ordered sequence of algebraic equations with
coefficients depending on both model parameters and previous roots in the
sequence. Briefly one finds first a quintic, as derived in the main body
of the paper. If the root with largest real part is real, the solution is
stable, if a complex pair, unstable. That root enters into coefficients of
the next polynomial, of seventh degree.  Its derivation is more involved
and hence confined to an appendix. A complex root pair of that polynomial
explains the Great Moderation of Keen's model.  Derivation of a final
polynomial depending on all the preceding is briefly sketched. In total
nine unique exponents result, which can be thought of as a nonlinear basis
set much as \cite{Cvitanovic:zeta} speaks of zeta functions as the
nonlinear equivalent of a Fourier transform.

Existence of such an asymptotic solution (that it has a {\sl finite}
spanning basis set of exponents) shows that long term trajectories are not
chaotic but rather entirely deterministic. The only uncertainty in the
solution arises instead in the preliminary adjustment phase, of about a
decade, where model conditions in the initial nine-dimensional parameter
space collapse onto a lower dimensional attractor. And even this latter
uncertainty is in principle amenable to resolution, requiring
approximation of the associated projection operator.

This study aims to identify those parameters in an economy that play the
greatest role in precipitating the endogenous instability that Minsky so
presciently foresaw or, alternately expressed, those parameters that, if
altered, could most easily mitigate the potential for instability.  A
careful delineation of the sensitivity of this economic model to the
various governing parameters reveals that just a few are key.  The most
critical are capital-output ratio, the return to capital, and the interest rate.

From the standpoint of policy, the capital-output ratio is an economic
factor hardly amenable to intervention in any practical way, but rather
reflects the complex interaction of diverse factors such as the overall
degree and nature of industrialization, of technological adoption, of
financialization of the economy, and so on. But the return to capital is
more immediately affected by, among other things, tax policy and the
bargaining power (or lack thereof) of unions, and the interest rate is of
course a direct function of monetary policy.

However the model explored in this first work is confined to investment,
the debt service for which is paid from the revenue stream generated by
the capital asset. Far more damaging is what Minsky identified as the
inevitable secular trend in good economic times towards increased ``Ponzi
finance,'' where investment in already existing assets fuels a bubble, and
debt service requires either further borrowing or else sell-off of the
asset in a (hoped for) rising market. Work that explores the latter
direction is reported in the very recent paper of
\cite{Grasselli}, who determine equilibria and the local stability of both
a third-order model by Keen and then the authors' extension to a
fourth-order system, which incorporates dynamical evolution of Ponzi
finance through the economic cycle. A key finding is that Ponzi financing
destabilizes what were otherwise stable equilibria of the system.

\section{Model Introduction}
\label{sect:model}

The model considered here is a system taken from \cite{Keen_arewe}, but
see also the detailed discussion and development of essentially the same
model in \cite{Keen:jebo}\footnote{But note the following
  errata in \cite{Keen:jebo}: $\tau_V$ and $\tau_L$ are undefined, though
  Keen intends to follow the definitions given earlier in
  \cite{Keen_arewe}, and (1.6), which sums the Table 1 entries, lacks the
  entries $j$ in the equations stated for $F_D$ and $F_L$, although these
  appear correctly in the subsequent (1.7).}:
\begin{eqnarray}
  \frac{d B_C}{dt} &=& \frac{F_L}{\tau_{RL}(\pi_r)} - \frac{B_C}{\tau_{LC}(\pi_r)}
\label{eq:sys1} \\
  \frac{d B_{PL}}{dt} &=& r_L\, F_L - r_D\, F_D - r_D\, W_D - \frac{B_{PL}}{\tau_B}\\
  \frac{d F_L}{dt}  &=& \frac{B_C}{\tau_{LC}(\pi_r)} - \frac{F_L}{\tau_{RL}(\pi_r)} + P_C\, Y_r\, Inv(\pi_r)\\ 
  \frac{d F_D}{dt}  &=& r_D\, F_D - r_L\, F_L + \frac{B_C}{\tau_{LC}(\pi_r)} - \frac{F_L}{\tau_{RL}(\pi_r)}
           + \frac{B_{PL}}{\tau_B }
             + \frac{W_D}{\tau_W}- \frac{Y_r\, W}{a(t)} 
    + P_C\, Y_r\, Inv(\pi_r)\label{eq:fd}\\ 
  \frac{d W }{dt}   &=& \left (Ph(\lambda) + \omega\, (g(\pi_r)-(\alpha+\beta))
         - \frac{1}{\tau_{Pc}} \, \left (1 - \frac{W}{a(t)\, (1-s)\, P_C} \right ) \right )\, W \\  
  \frac{d P_C}{dt}  &=& - \frac{1}{\tau_{Pc}}\, \left (P_C - \frac{W}{a(t)\, (1-s)}\right ) \label{eq:pc}\\  
  \frac{d K_r}{dt}  &=& g(\pi_r)\, K_r\\       
  \frac{d \lambda}{dt} &=& (g(\pi_r)-(\alpha+\beta))\, \lambda \label{eq:sys8}
\end{eqnarray}
where $a(t) = a_0 \, \exp(\alpha \, t)$ models the growth of productivity
over time.  Population growth is modeled as $N_0 \exp(\beta \, t)$ (the
effect of which is incorporated implicitly in the system above via the
growth rate $\beta$, in contrast to the explicit dependence on $a(t)$).
The rate of profit is
\be
  \pi_r  = \frac{1}{v\, P_C\, Y_r}\,  \left (P_C\, Y_r - W\, Y_r/a(t) 
 - \left (r_L\, F_L - r_D\, F_D \right )\right )\, .\label{eq:profit}
\ee
We relate capital to output via $ Y_r = K_r/v$.
The rate of economic growth is given by $ g =
Inv(\pi_r)/v - \delta$.  
Auxiliary functions defined in terms of a generalized exponential are
\begin{eqnarray*}
  Inv(\pi_r)   &=& G(\pi_r,1/25,1/25,2,0)\\
  Ph(\lambda)     &=& G(\lambda,96/100,0,2,-1/25)\\
  \tau_{RL}(\pi_r) &=& G(\pi_r,3/100,10,100,3)\\
  \tau_{LC}(\pi_r) &=& G(\pi_r,3/100,2,-50,1/2)
\end{eqnarray*}
where 
\[ G(x,x_\nu,y_\nu,s,m) = (y_\nu - m)\, \exp\left (s\, \left (x-x_\nu
\right )/\left (y_\nu-m\right )\right ) + m
\, .
\] 

This system differs from Keen only in that the evolution equation for
worker deposits, $W_D$, has been eliminated since, by an accounting
identity, the sum of $ W_D(t) + F_D(t) + B_{PL}(t) - F_L(t) \equiv cst $
is invariant in time and so $W_D$ can be always computed in terms of the
remaining variables. Values for the parameters are chosen as stated in
Table~\ref{table:standard} and initial conditions in
Table~\ref{table:IC}. The latter are supplemented with $ W_D(0) = 13$ and
hence $cst = 12$.

\begin{table}
\begin{caption}{Standard parameters for the model specified in (\ref{eq:sys1}
- \ref{eq:sys8})}
\label{table:standard}
\end{caption}
\begin{center}
\begin{tabular}{cccc}
\noalign{\smallskip}
\hline\noalign{\smallskip}
 $\alpha = 0.015$ &  $\beta  = 0.02$ &   $\delta = 0.01$&$\omega = 0.1$ \\  
 $\tau_B =  1 $   & $\tau_W =1/26$ &  $\tau_{Pc} =  1 $   & $ s   = 0.27 $ \\
 $ a_0   = 1$ &  $ v=3$ & $ r_L   = 0.05$ & $r_D   = 0.01$ \\ 
\noalign{\smallskip}\hline
\end{tabular}
\end{center}
\end{table}

\begin{table}
\begin{caption}{Initial conditions for (\ref{eq:sys1} - \ref{eq:sys8})}
\label{table:IC}
\end{caption}
\begin{center}
\begin{tabular}{cccc}
\noalign{\smallskip}
\hline\noalign{\smallskip}
  $B_C(0) = 12$  & $B_{PL}(0)=  5$ & $F_L(0) =100$ & $F_D(0) = 70 $ \\
  $W(0)  =  1$  &  $P_C(0) =  1$ & $ K_r(0) =900$&  $ \lambda(0)= 1$\\
\noalign{\smallskip}\hline
\end{tabular}
\end{center}
\end{table}

\begin{figure}
\begin{center} 
\epsfxsize=4in{\epsfbox{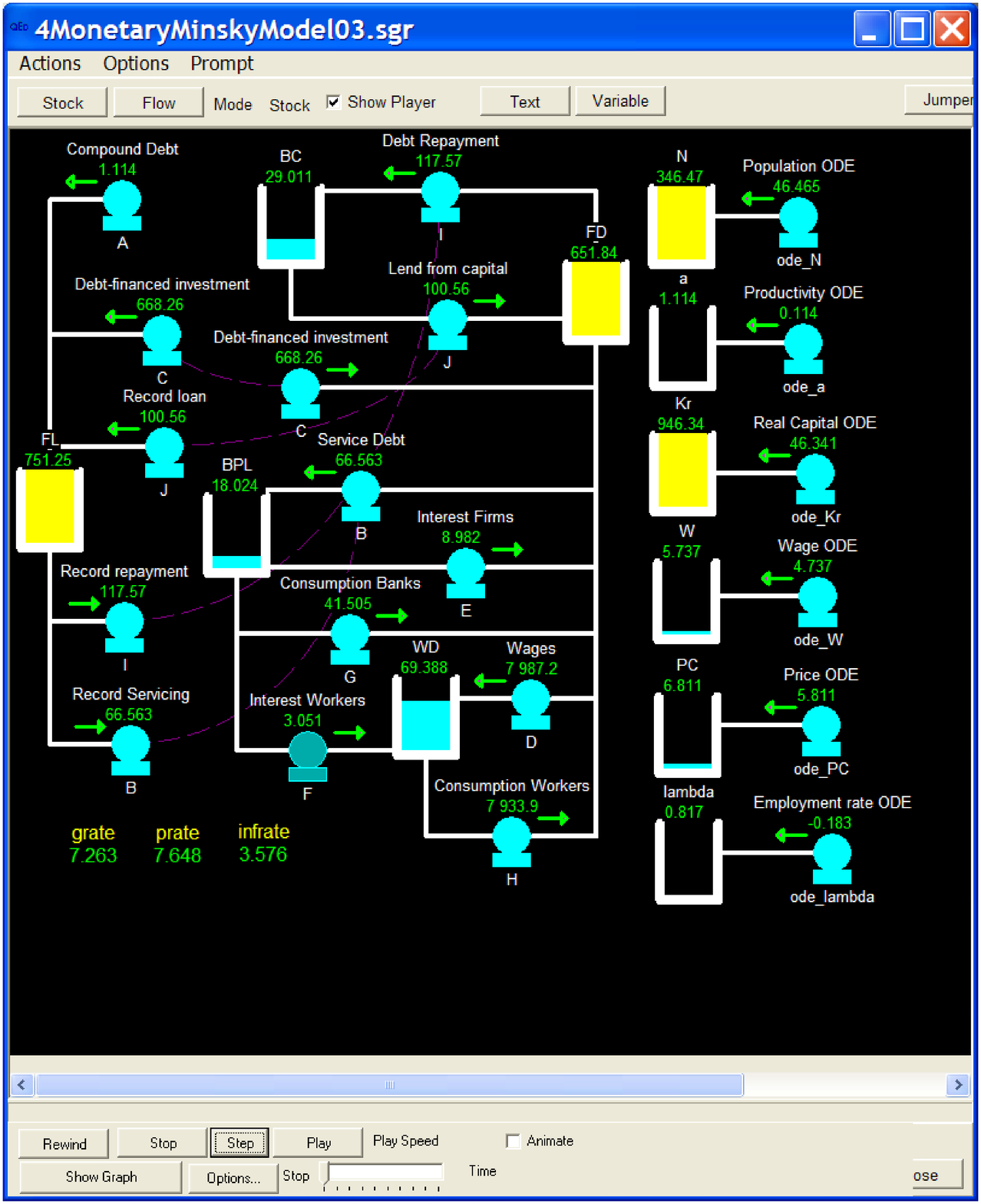}}
\begin{caption}{A Phillips diagram of (\ref{eq:sys1} -- \ref{eq:sys8})
prepared using the QED software package developed by an anonymous collaborator
of Steve Keen.}
\label{fig:phillips}
\end{caption} 
\end{center} 
\end{figure}

Briefly, $B_C$ is bank capital, $B_{PL}$ the flow of funds in and out of
the bank in e.g., interest payments on deposits and loans, $F_L$ monies
borrowed by a firm for investment, $F_D$ firm deposits at the bank, $W$
wages paid to workers, $P_C$ the price of goods, $K_r$ capital, and
$\lambda$ the employment rate.  A block ``Phillips'' diagram that
represents the interaction of these variables is shown in
Figure~\ref{fig:phillips}.

The central assumption of the model is the apostate view of
\cite{schumpeter} on endogenous expansion of credit money by the banking
system.  As noted in \cite{Fontana}, the modern monetary theory of
production has two branches; the Circuitist school --- reflected in this
model in the specific appeal to the three agent model of \cite{Graziani}:
a buyer, a seller, and a bank --- and the post-Keynesian school. The focus
of the latter on liquidity preference arising from the zero elasticity of
production and substitution of money is not part of the model but one can
imagine various means of its incorporation. Moreover the model is about
the growth of credit money, not fiat money of a sovereign government {\sl
  per se}, although the existence of the latter is implicit in the tacit
acceptance of currency itself. But here, too, a suitable generalization is
readily envisioned, with the central distinction {\sl contra} Graziani, of
the asymmetry of only two agents --- government and non-government ---
giving rise to the point first noted by \cite{Lerner} of the irrelevance
of government ``debt''. The model {\em is} about the dynamics of debt,
whose salience at the present time is apparent. Precisely because of the
ubiquity of instability, perhaps the most interesting direction in which
one might extend the model would be to include a significant role for
equity-based, rather than debt-based, financing, the stabilizing influence
of which has long been argued by Michael Hudson. 

Perhaps the most important addition to Keen's model in light of the
post-industrial financialization of the economy would be a store of value
modeled by a lumped parameter asset proxy representing equities and real
estate, introducing the competitive relation between asset pricing and
commodity and wage pricing.

Other significant features include increased (``speculative'') borrowing
over and above profits by firms during booms and diminished (``hedge'')
borrowing below the prevailing profit level during slumps, and upward wage
pressure in a tight labor market. Note finally that in this simple model
there are no explicit fixed (depreciating) capital assets, only
deposits. And workers live always within their means; they do not take out
loans.

While greatly simplified in many respects, Keen's model is nonetheless an
extremely valuable paradigm for understanding endogenous money creation as
well as the intrinsic instability of a debt-based economy. This is all the
more so true given that, as this paper demonstrates, one can completely
characterize all the essential dynamics in analytic terms and hence
dissect all the parametric dependences of the model with no need of time
stepping numerous particular cases, looking for common patterns by heuristic
means.

\section{Analysis}
\label{sect:anal}

Initially there are eight unknowns governed by eight equations; the coefficients of the functions
\[
[ B_C , B_{PL}, F_L, F_D, W, P_C, K_r, \lambda ] \, .
\]
As noted, we eliminate $W_D$ from the identity that $F_L - F_D - B_{PL} -
W_D$ is constant. The value of this initial constant is subdominant
relative to assumed exponentially growing solutions and so is dropped from
the system, that is, the so-called ``leading order'' balance imposes that
the four exponential terms sum to {\sl zero}.\footnote{All the
mathematical arguments developed here are standard. There are numerous
available texts, the author prefers \cite{BO}} Absent
that initial constant, the resulting system is homogeneous in the
unknowns. To fix a solution of this nonlinear system requires that one
then identify two of the unknowns as fiducial parameters in terms of which
the remainder are given.

Note the force of this observation. We start off the economy by having to
specify nine distinct, unrelated, quantities (including $W_D(0)$). The
initial firm loan, $F_L(0)$, has nothing to do with the starting wage,
$W(0)$, neither constrains the initial price level, $P_C(0)$, and so
on. And yet after a short time, effectively regardless of how we start,
knowledge of only {\sl two} of the economic variables suffices to
determine the value of the others; nonlinearity {\sl forces} particular
relations among these. This is in stark contrast with a linear system of
eighth order. For the latter one must at all times observe the current
values of all eight variables to uniquely specify the state of the
system.\footnote{This should be qualified by saying ``almost always'' as
there are technical exceptions, such as 
regular and irregular singular points, where fewer than
eight values could still fix the solution uniquely.}

This partition is not unique, but
one convenient choice is to identify $B_C^{(0)}$ and $K_r^{(0)}$ as the
pair. With this choice the appropriate form for the variables is found to
be
\begin{eqnarray}
 B_C &\sim& B_C^{(0)}\, \exp(\mu\, t) \label{eq:bceq} \\
 B_{PL} &\sim& B_{PL}^{(0)} \, B_C^{(0)}\, \exp(\mu\, t) \\
 F_L &\sim& F_L^{(0)} \, B_C^{(0)} \, \exp(\mu\, t)\label{eq:fleq}\\
 F_D &\sim& F_D^{(0)} \, B_C^{(0)}\, \exp(\mu\, t)\label{eq:fdeq}\\
 W &\sim& W^{(0)} \, B_C^{(0)}/K_r^{(0)}\,  \exp((\mu - \beta)\, t) \label{eq:weq}\\
 P_C &\sim& P_C^{(0)} \, B_C^{(0)}/K_r^{(0)} \, \exp ((\mu - \beta -  \alpha)\, t) \label{eq:price}\\
 K_r &\sim& K_r^{(0)} \, \exp((\alpha+\beta) t) \label{eq:Kreq}\\
 \lambda &\sim& \lambda^{(0)} = 24/25 + 1/50 \, \log(1+25\, \alpha) 
\label{eq:lameq}\, .
\end{eqnarray}
Note that $\lambda$, the employment rate, is asymptotically constant and
independent of initial conditions, fixed solely by the rate of change of
productivity, $\alpha$.  Additionally, $g$ tends to the value $\alpha +
\beta$ so that the equation for $ K_r$ is immediately solved without
reference to the unknown $\mu$, or the other variables.  
For this solution the debt
ratio, $100\, F_L / (P_C \,Y_r)$, is asymptotically constant.

One can give a partial intuitive justification for this scaling. The
variable $K_r$ measures capital but, by virtue of a fixed capital to
output ratio, this is a direct measure also of real output of goods. By
contrast $B_C$ is strictly a measure of credit money. As the price of
goods is variable, there must then be {\sl at least} one added scale
parameter in the problem related to finance. The surprise is that one
alone suffices.  Note that the temporal output of real goods is completely
decoupled from the behavior of money in this debt-based economy. Growth in
production of goods in this asymptotic regime is rigidly locked to
population growth and gains in productivity {\sl regardless} of the
financial state, i.e.\ whether a stable or unstable growing solution.

The forms above do not constitute exact solutions of the equations, only
leading order approximations, that is, there is a series of small
corrections.  But the exponential terms dominate as $t \to \infty$.  It
should be noted that there is no theory for the form of an asymptotic
expansion except for a very restricted class of linear ordinary
differential equations. Even for a second order linear problem, but at an
irregular singular point, one can in general only rely upon experience and
intuition to say nothing, as here, of a {\em nonlinear} ordinary
differential equation, and ninth order at that. However, it is known that
if a solution {\em has} an asymptotic expansion, then the form of the
latter is unique.\footnote{Note the freedom to choose fiducial variables
  simply defines an equivalent family of solutions; this does not vitiate
  the statement.} In nearly every case in practice, the correctness of the
proposed form has to be established by showing the expansion is consistent
order-by-order up to a given level of truncation. That exercise is carried
out in Appendix B, where explicit construction of higher order terms
vindicates the leading order forms given here and, further, the complete
basis of a general solution is established by showing that this expansion
has the requisite nine free parameters (reincorporating the equation for
$W_D$). This is to be contrasted with the common case of so-called
``singular'' asymptotic solutions that have fewer free parameters than the
order of the governing equation(s). In the end, the acid test of any
proposed asymptotic expansion is that it matches the computed solution
with an error term that vanishes at the correct rate in the relevant limit
(here $t \to \infty$).  The various figures that follow demonstrate this
conclusively.

What we obtain from this exercise is a solution which, carried out to four
terms for each of the independent variables, gives a reasonably accurate
solution commencing at say, $t=15$, and which becomes exponentially more
accurate for increasing $t$. The logical complement to this is a solution 
exact at $t=0$, hence taking on board the initial conditions, and that 
gives also a tolerably accurate result at $t=15$. The two solutions,
blended together using the formal method of matched asymptotic expansions,
then constitute a global solution for Keen's model. All the significant
dynamics of Keen's model are contained in the large time solution. The main 
utility of a short time solution is to characterize the space of initial 
conditions.

The key that permits the large time solution, and motivates the ansatz above, is
that the nonlinear terms are all simple ratios or products and so the
various exponent dependences can add in an appropriate fashion such that a
set of terms in a given equation share a common exponent while other
terms, if any, have a lesser exponent. This is an application of the
well known method of ``dominant balance''.

Degeneracy of this set enters from comparison of equations for $d W/dt$
and $d P_C/dt$. A nontrivial solution for the pair requires that we take
$Ph(\lambda) = \alpha$, which in turns defines the constant value of
$\lambda^{(0)}$, but then this pair of equations is replaced by just the
one for $dP_C/dt$. The other option is to choose $s=1$ but this yields a
vacuous solution. This reduces the problem to an overdetermined system of
six (linear!) equations in the five unknowns $[B_{PL}^{(0)}, F_L^{(0)},
  F_D^{(0)}, W^{(0)}, P_C^{(0)} ]$: (\ref{eq:sys1} -- \ref{eq:fd}),
(\ref{eq:pc}), and (\ref{eq:profit}).  However, $\mu$ enters as a free
parameter in the coefficient matrix and hence it is possible to obtain a
consistent system if $\mu$ is chosen so that the determinant of the
augmented $6 \times 6$ matrix vanishes. This yields a quintic polynomial
for $\mu$ (only quintic since (\ref{eq:profit}), which closes the system, is a
kinematic consistency condition, rather than a dynamic constraint).

In the instance that $\mu$ is real, a realizable model requires that the
solution for each of the five unknowns be non-negative. That restriction 
defines limits on model parameters, one example of which appears in a later
figure.

Not including the sixteen parameters that enter into the generalized
exponential functions, there are ten other parameters whose values
determine the roots for $\mu$. The expressions for the coefficients are
too complicated to report here but are readily programmed via symbolic
means so that one can then numerically explore the signature of
roots over the entire ten-dimensional space. (The generated code runs
to about 1200 lines.)

Before turning to that exploration, however, it is conceptually helpful to
consider the single parameter $s$, the return to capital, set in the model
at $0.27$ in accord with historical norms. Understanding its influence
provides a useful paradigm for understanding the more general parametric
dependence of the model. With the other nine parameters fixed at the values
in Table~\ref{table:IC}, the following polynomial for $\mu$ results
\begin{eqnarray}
\label{eq:sbif}
0.5583&\, (s - 1)\, 
\mu^5 + (16.2171\, s - 15.7665)\, \mu^4 + (45.9571 \, s - 33.31195) \, \mu^3 
 \nonumber \\
 &+ (45.4431 \, s - 20.6586) \, \mu^2 + (15.1452 s - 2.6459)\, \mu - 0.080144
\, .
\end{eqnarray}
At the standard value of $s$, there are three real negative roots of
$\mu$. These violate the assumption of a growing
solution. The remaining two are the complex pair $ 0.06987723 \pm
0.04598378\, i$. As $s$ is increased, the imaginary component
approaches zero and at $s_{crit} = 0.2891574$, a double real root
appears. Above this bifurcation point, there are two positive real roots.
The root of largest real value is hereafter denoted $\mu_0$ (and the
one with 
positive imaginary part in the case of a complex pair).

\subsection{Unstable solution}
\label{sect:unstable}

\begin{figure}
\begin{center} 
\epsfxsize=3.5in{\epsfbox{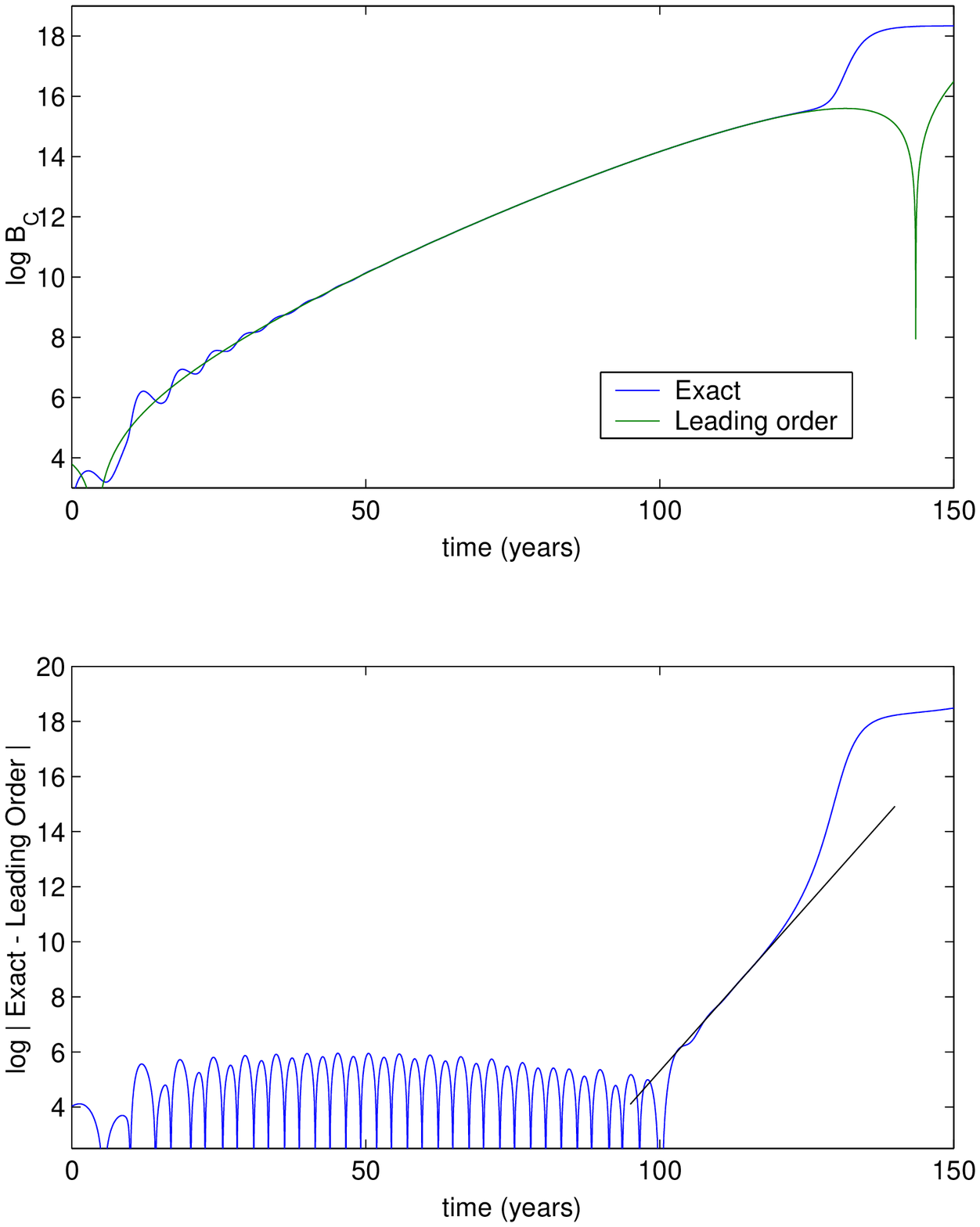}}
\begin{caption}{{\it Top panel}: comparison of the leading order prediction for $B_C$ and the actual computed solution. {\it Bottom panel}: the logarithm of the difference of the two curves, showing the onset of instability commencing at about $t=100$}
\label{fig:unstable}
\end{caption} 
\end{center} 
\end{figure}

Looking first to the complex case, which corresponds to an unstable
financial system, we take $s = 0.285$, fairly close to the bifurcation
point and on the unstable side. Here $ \mu_0 = 0.08146881 + 0.02251608 \,
i$.  To compare (\ref{eq:bceq}) to numerical simulation requires that we
determine the unknown phase and amplitude. While these are in principle
unique functions of the initial conditions, that relation is not immediate
and so instead the values are here fixed numerically by simply matching
the long time behavior.  This exercise gives a leading order estimate of
\be\label{eq:bca}
 B_C \sim |B_C^{(0)}| \, \exp( \Re(\mu_0)\, t) \, \sin ( \Im(\mu_0) \, t + \phi )
\ee
where $|B_C^{(0)}| = 494.24171$ and $\phi = -0.090296116$.
Figure~\ref{fig:unstable} shows an initial period of adjustment of the exact solution, with
oscillation of decreasing period about the leading order result. In the
lower panel one sees that this vacillation quickly assumes a well defined
period, the value of which can be found by carrying analysis to next
order, for which see Appendix B.

Notice the transparent onset of instability in the lower panel, at about
$t=100$, its interval of small (relative) amplitude indicated by the solid
line.  The rate of growth of this initial transient (the slope of the
straight line) could be determined from suitable analysis but this is
aside from the main thrust of the paper. Rather we remain focused on the
goal of explicating all possible limiting behaviors and so consider only
late stage evolution of the solution, characterized in \S
\ref{sect:collapse}.

Perhaps the most significant feature to note is the sharp dip of the
leading order curve in the top panel at about $t=143$. This reflects the
zero crossing of the sine function in (\ref{eq:bca}). There is a second
dip at $t=3$ but this is in the interval of initial adjustment as the
solution evolves from the given initial conditions onto the ``manifold''
defined by the exponential solution.  {\sl Of necessity}, a complex
exponential solution invalidates itself; even discounting that predicted
negative values for the various variables are meaningless, that the terms
assumed largest in the equations become arbitrarily small is 
inconsistent. While we are not yet, based on analysis, able to say what
becomes of the solution subsequently, we nonetheless can give a definite
estimate for the breakdown time, namely $ T = \pi / \Im(\mu_0)$, here
$139.5$ years.  It is fundamentally this half-period dictated by the root
of the quintic that sets the basic duration of economic growth.

Late in the cycle there is the noted onset of instability, whose growth rate is
much faster than $\Re(\mu_0)$, and this precipitates the ultimate
(nonlinear) collapse.  As remarked above, that fatal instability cannot
occur early in the cycle; the conditions for its growth are unfavorable
then. So we may accept $T$ as a leading order approximation of the
duration of the economic growth cycle (or equivalently, the time until
collapse), with modest correction of that based on a detailed analysis of
the transition to collapse, briefly limned in \S \ref{sect:collapse}.

\begin{figure}
\begin{center} 
\epsfxsize=3.5in{\epsfbox{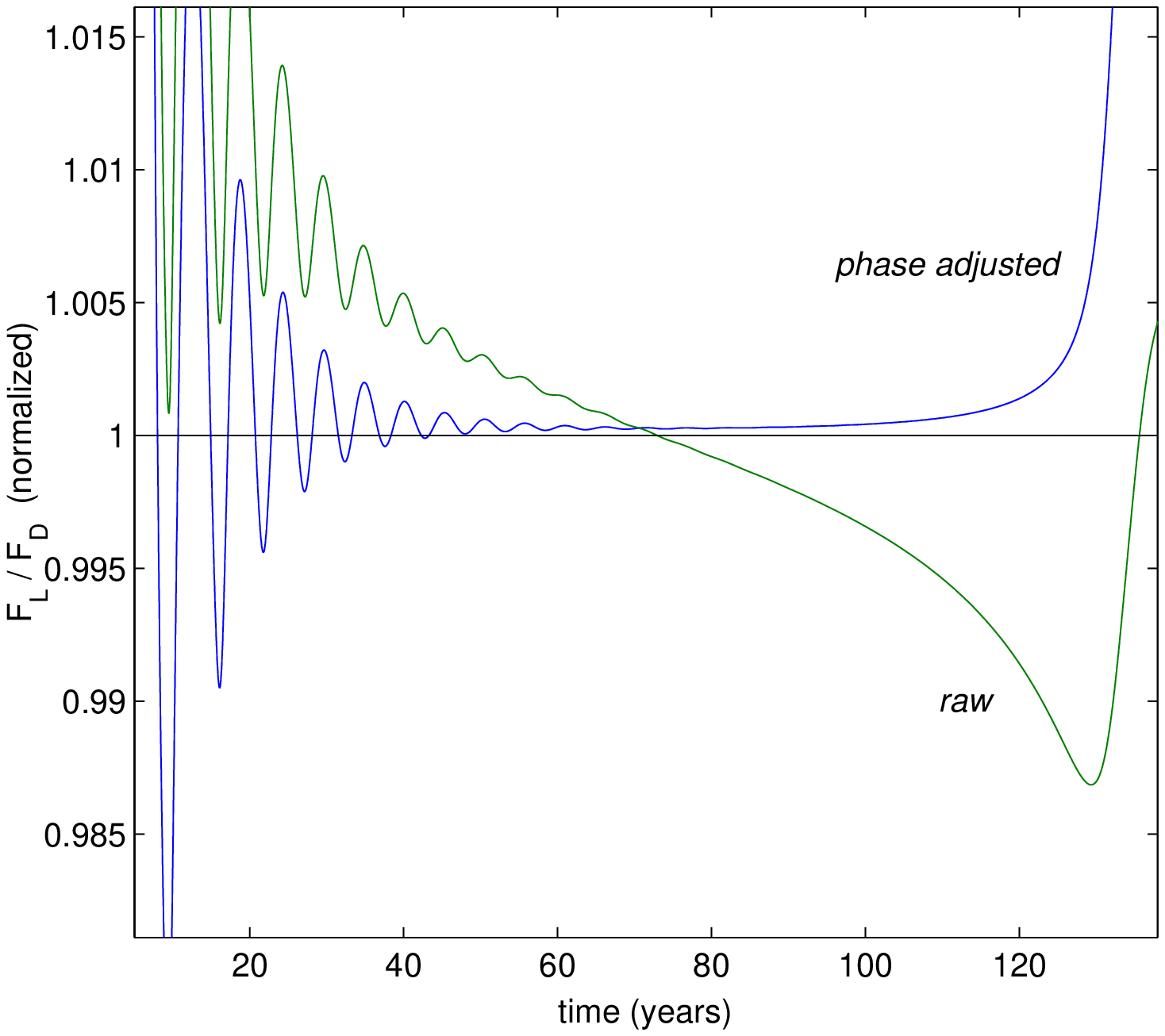}}
\begin{caption}{A second test of the asymptotic theory: the ratio
should approach unity.}
\label{fig:flfd}
\end{caption} 
\end{center} 
\end{figure}

Finally, while (\ref{eq:bca}) does match the computation very well, that
comparison rests upon two empirical parameters. A more exacting, parameter-free
test is desirable. For this purpose, (\ref{eq:fleq}-\ref{eq:fdeq}) suggest that one
compute the ratio of $F_L/F_D$, as this should approach a constant. That
value of that constant is a specific prediction
from the solution of the overdetermined linear set noted in 
the first part of \S \ref{sect:anal}. For $s=0.285$, the exact values of the
relevant coefficients are found to be 
\[
F_L^{(0)} =  15.935787 + 0.32483068 \, i\qquad
F_D^{(0)} =  15.050902 + 0.23250923 \, i \, .
\]
Converting these complex results to equivalent real-valued solutions gives
the ratio of the moduli of these two complex numbers as the desired
constant, $1.058886$.  The ratio of $F_L/F_D$, divided by that prediction,
is plotted as the ``raw'' curve in Figure~\ref{fig:flfd}. That result is
not seen to approach unity.  The reason is that the complex results {\sl
  also} determine a phase shift of $F_L$ with respect to $F_D$, as
reflected in the arguments of their respective sine functions following
conversion to real form. For the given parameters, deposits at a given
time should be compared to loans slightly more than two and a half months
earlier; deposits lag loans. That result (``phase adjusted'') conforms to
expectation extremely well; the ratio levels off to within a few parts in
$10^4$ of unity. (The later sharp upsweep of the curve marks the
transition to collapse.)  The theory is vindicated.

This phase correction points up a needed revision in the proposed estimate
of $T$. Really one should solve the augmented $6 \times 6$ system, as was
done for the entries above, and the phases of the five fields relative to
$ B_C$ determined. The complete results of this exercise for the unstable
case here are given in Table~\ref{table:ivp}, expressed in real form.  The
nearest zero crossing over this full set marks a tighter upper bound on
the duration of growth, always less than half a period. Here (and in most
cases) that restriction comes from the phase shift for wages, $ W$, and the
constrained period is $T^* = 125.74$ years. For smaller values of $T$,
this restriction can become a significant relative decrease. Nonetheless,
for simplicity, in the remainder of the paper we continue to cite the
formula for the elementary estimate, $T$, as an explicit representation
for the phase correction is not immediately revealing.\footnote{See
\ref{sect:appendixA} for some detail.}

\begin{table}
\begin{caption}{Scale amplitudes and phases for
the unstable (complex) version of the asymptotic solution 
in (\ref{eq:bceq} - \ref{eq:lameq}) that matches the model
with parameters as given in Table~\ref{table:standard} except
for $s = 0.285$}
\label{table:ivp}
\end{caption}
\begin{center}
\begin{tabular}{ccc}
\noalign{\smallskip}
\hline\noalign{\smallskip}
  variable   &  amplitude    &   phase (years) \\
\noalign{\smallskip}\hline\noalign{\smallskip}
$B_{PL}^{(0)}$ &  $0.5949050 $ & $-0.0277274$ \\
$F_D^{(0)}$    & $15.0526978 $ & $ 0.6862392$ \\
$F_L^{(0)}$    & $15.9390976 $ & $ 0.9054319$ \\
$W^{(0)}$      & $23.8111979 $ & $13.7830358$ \\
$P_C^{(0)}$    & $31.8161573 $ & $12.8275901$ \\
\noalign{\smallskip}\hline
\end{tabular}
\end{center}
\end{table}

To reiterate an earlier observation, we might have launched this model
economy with a million dollars in firm loans outstanding and one dollar in
the firm's deposit account. Or the reverse. And yet after a modest
interval, that ratio of firm loans to firm deposits (suitably lagged)
approaches parity ($1.0588..$), and without regard to any other initial
variable, such as the interest rate, or worker salary, or the employment
rate! The more general import of this is the suggestion that the most
useful economic indices will generally be ratios of time-lagged aggregate
quantities. Which ratios are of most diagnostic value will depend on the
quality and frequency of available datasets.  Optimal lags must be
discovered empirically.

\subsection{Stable solution}
\label{sect:stable}

\begin{figure}
\begin{center} 
\epsfxsize=3.5in{\epsfbox{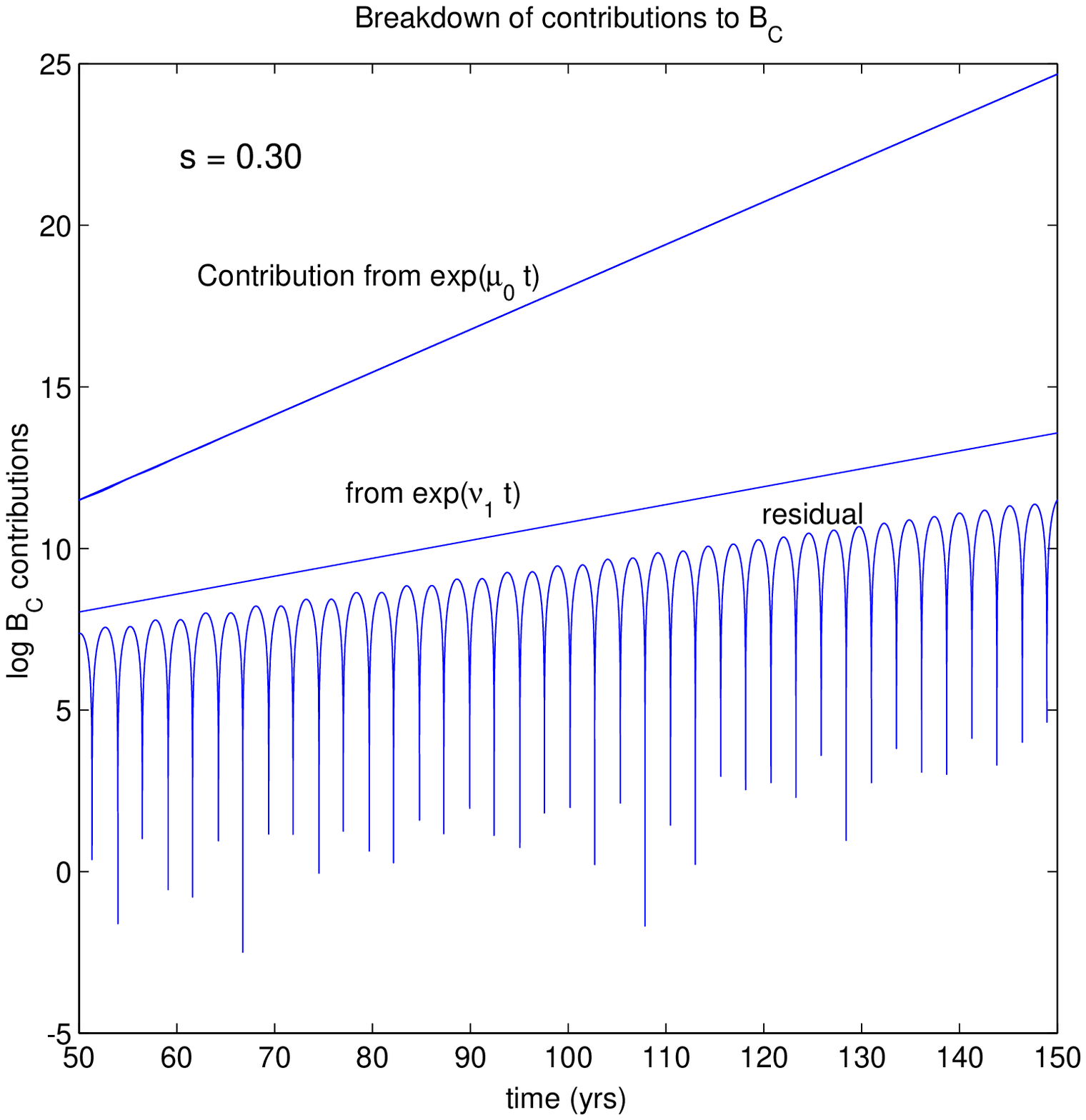}}

\begin{caption}{Dissection of the exact solution in terms of (\ref{eq:bcb}).}
\label{fig:stable}
\end{caption} 
\end{center} 
\end{figure}

\begin{figure}
\begin{center} 
\epsfxsize=3.5in{\epsfbox{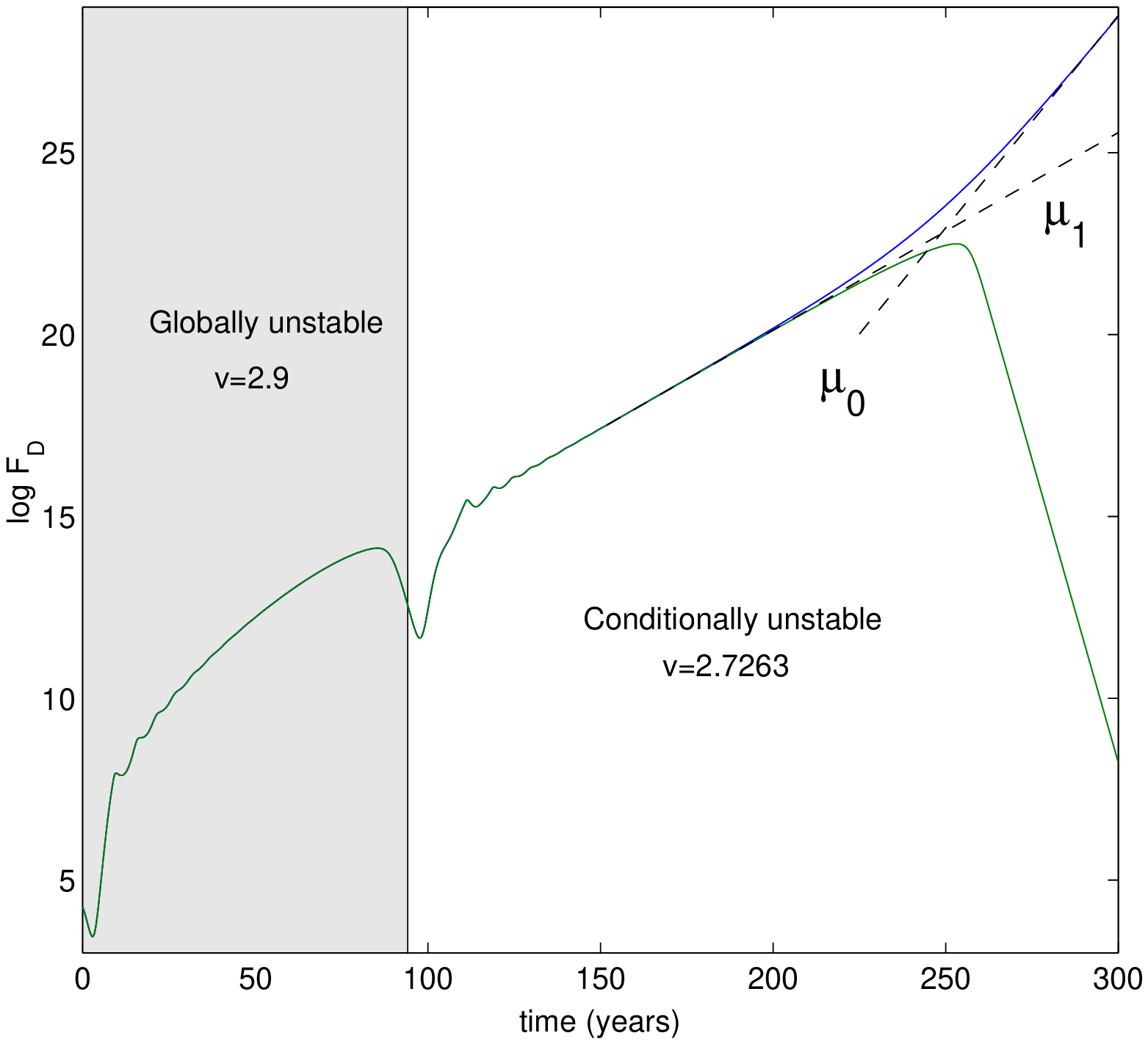}}

\begin{caption}{A slight perturbation of initial conditions to show that 
either immediate collapse or stable long term growth is possible.
}
\label{fig:bistab}
\end{caption} 
\end{center} 
\end{figure}

Turning to $s > s_{crit}$, we take $s = 0.3$ and find 
the largest root is $\mu_0 = 0.13171713$. We can here anticipate a
correction to the leading order form
\be\label{eq:bcb}
 B_C \sim B_C^{(0)} \, \exp( \mu_0 \, t) + B_C^{(1)} \, \exp( \nu_1 \, t) \, .
\ee
For Figure~\ref{fig:stable} the coefficients $B_C^{(0,1)}$ are determined by a
least squares fit.  Note the oscillatory residual, once these two are
removed, which is {\sl also} exponentially growing, but is subdominant to
the other two. Empirically this solution is stable; growth persists
indefinitely. To prove its linear stability would be a tedious exercise
but also a weaker result than desired. Linear stability shows only that the 
solution is stable to infinitesimal disturbances where this solution has
in fact finite amplitude stability; it recovers from observable disruptions.
A yet stronger result would be a demonstration of ``global stability'',
meaning that this solution is the universal long term outcome for all 
possible initial economic states. But that is easily shown to be false 
by counterexample as indicated in Figure~\ref{fig:bistab}, where we 
introduce a bifurcation induced now by $v$, the capital-output ratio,
rather than $s$ as above, and track the behavior of $F_D$.

The initial computation uses $v=2.9$, which is still in the unstable
complex exponential regime. The computation is then halted at
$t_1=94.16736$ and again at $t_2=94.16748$. Each of these two interrupted
computations is then continued but with $v = 2.7263$, for which $\mu_0$ is
real and hence a stable growing solution exists as that above.

But note that while the first computation carrying on from $t_1$ (upper
curve) does eventually approach the stable growing solution with predicted
growth rate $\mu_0 = 0.1167$, the second one commencing from $t_2$ (lower
curve) ultimately ends in collapse. We can infer that there is a
particular value $t^*$ in the open interval $(t_1,t_2)$ that does neither,
but rather continues along the indicated dashed line with slope $\mu_1 =
0.05428$. This is the second positive root of the quintic; it is an
unstable solution. The set of all possible initial conditions that yield
this second solution constitutes a border in the eight-dimensional space
of initial conditions; dividing solutions that grow indefinitely at rate
$\mu_0$ from those that collapse.\footnote{This scenario assumes $\mu_0 >
  \mu_1 > 0$. A solution on the separatrix for the case of real roots with
  $\mu_0 > 0$ and $\mu_1 < 0$ would exhibit some other time dependence,
  but such a case does not appear to arise for realistic parameter values.}
As the present computation hints, conditions that yield the latter are
uncommon.  In effect, only when the initial state is such that things are
already headed towards collapse is that scenario likely to continue.

\subsection{Collapse and predictability}
\label{sect:collapse}

The late stage evolution of the solution heading for a collapse is as follows.
First note that $\pi_r$ diverges as a negative definite exponentially
growing function and hence the functions 
$Inv$, $\tau_{RL}$ and $\tau_{LC}$ quickly saturate at 
$0$, $3$, and $\infty$ respectively. Also $\lambda$ is a positive definite
exponentially decaying function and so quickly $Ph$ saturates at $-4/100$. 
We have then in addition that $ g = -\delta$.
(These simplifications are easily justified as self-consistent 
after the fact.)

It follows that 
\begin{eqnarray}
   B_C &\sim&  bc_0+bc_1 \, \exp(-t/\tau_{RL})\label{eq:crstart}\\
   F_L &\sim& fl_1 \, \exp(-t/\tau_{RL})\\
    Kr &\sim& kr_1 \, \exp(-\delta \, t)\\
   \lambda &\sim& \lambda_1 \, \exp(-(\alpha + \beta + \delta) \, t)
\end{eqnarray}
Next one solves for the pair $P_C$ and $W$. This pair is exactly soluble
on the substitutions noted above and yields
\begin{eqnarray}
 P_C  &\sim&  \exp((\exp(-(\alpha-Ph+\mu)\, t)\, c_1 - c_2)/(\alpha-Ph+\mu)+(1-t)/\tau_{Pc})\\
 W   &\sim& c_1 (s-1)\, a_0 \, \tau_{Pc} \, \exp((Ph-\mu)\, t) \, P_C(t)\label{eq:weq2}
\end{eqnarray}
where $\mu = \omega \, ( \alpha + \beta + \delta)$. Notice the exponential
within the exponential in the form for $ P_C$. These two results are easily
confirmed by using simulation results to plot
$ P_C/W \, \exp((Ph-\mu)\, t) $, which 
quickly saturates to a constant (and allows immediate determination of $c_1$).

This leaves one to solve for $F_D$ and $B_{PL}$. There are several
exponentials contributing to the solution for each but, for the parameters
as given in the standard model, all of these except the leading
contribution can be safely neglected leaving
\be
 F_D \sim fd_0 + fd_1\, \exp(-t/\tau_{RL}) \qquad
 B_{PL} \sim  bp_0 + bp_1 \, \exp(-t/\tau_{RL}) \label{eq:crend} \, .
\ee
It is an involved exercise to work out the relations among the various 
coefficients in these asymptotic forms but it is anyway instructive to note
that $bp_0 = cst/99$, where $cst$, recall, is the conserved quantity 
$F_L(t) - F_D(t) - B_{PL}(t) - W_D(t)$.

From these functional forms, one can derive the previously asserted
exponential divergence of $\pi_r$ and so the set constitutes a valid
asymptotic solution for the set of eight differential equations. Unlike
the previous solution however, this solution does not lead to any
characteristic polynomial whose coefficients depend on the parameters in
the problem and so it does not have an associated regime diagram like that in 
later Figure~\ref{fig:regime}. Rather, this solution exists for all parameter
settings. 

\begin{figure}
\begin{center} 
\epsfxsize=3.5in{\epsfbox{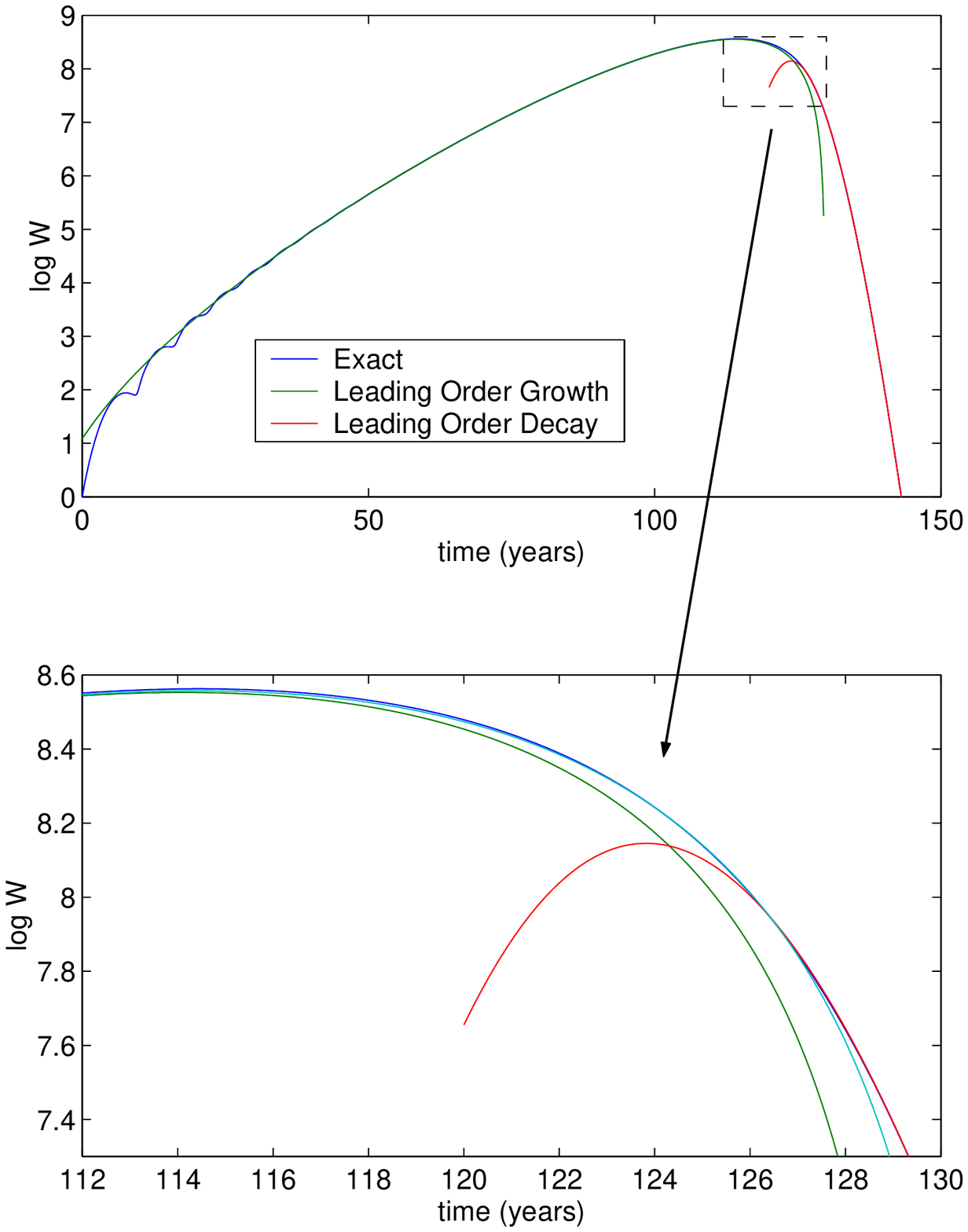}}

\begin{caption}{Prediction for $W$ in the terminal collapse.}
\label{fig:matching}
\end{caption} 
\end{center} 
\end{figure}

In Figure~\ref{fig:matching} the prediction from (\ref{eq:weq2}) (right
hand dashed curve) is compared to the exact solution for the same unstable
case as in \S \ref{sect:unstable}, where $ s = 0.285$. (As before, the
needed parameters are obtained from a numerical fit. These are $c_1 =
-1654.10346$ and $c_2 = -8.74399851$.) In addition the predicted
exponential growth from (\ref{eq:weq}) is plotted. It is seen that the two
results cover nearly the entire span of time. The narrow time interval of
about $[110,125]$ is the transition from one asymptotic form to the other.

This bridge is effected by the same instability denoted by the
straight line segment in Figure~\ref{fig:unstable}. In first
approximation, the slope of that line
emerges as the eigenvalue of a particular matrix. As an expedient
substitute for that calculation, the slope was determined {\it ex post
  facto} to be about $0.24$.  The sum of this growing transient $ c_3\, \exp(0.24 t)$,
added to the complex exponential solution, is plotted as a dash-dot line
in the lower panel (but not the upper) of Figure~\ref{fig:matching}.  One
observes that the sum intersects the second dashed curve (the decay solution)
tangentially. This is a simple graphical construction of what would emerge
from the method of matched asymptotic expansions. By means of that
formalism the amplitudes and phase, all here determined empirically, are
instead found deductively from the requirement that successive parts of
the solution be smoothly joined.

This view of the solution raises the issue of sensitivity to initial
conditions. Over most of the cycle, evolution of the two exponential
solutions, growing and decaying, is perfectly stable. Any variability in
time scale for collapse must therefore arise from either: (1) the initial
transient adjustment as the solution tries to lock on to the growing
solution, or (2) the timing of the trigger for the second transient
parasitic solution that acts as the bridge from one to the other
exponential.  

\begin{figure}
\begin{center} 
\epsfxsize=4in{\epsfbox{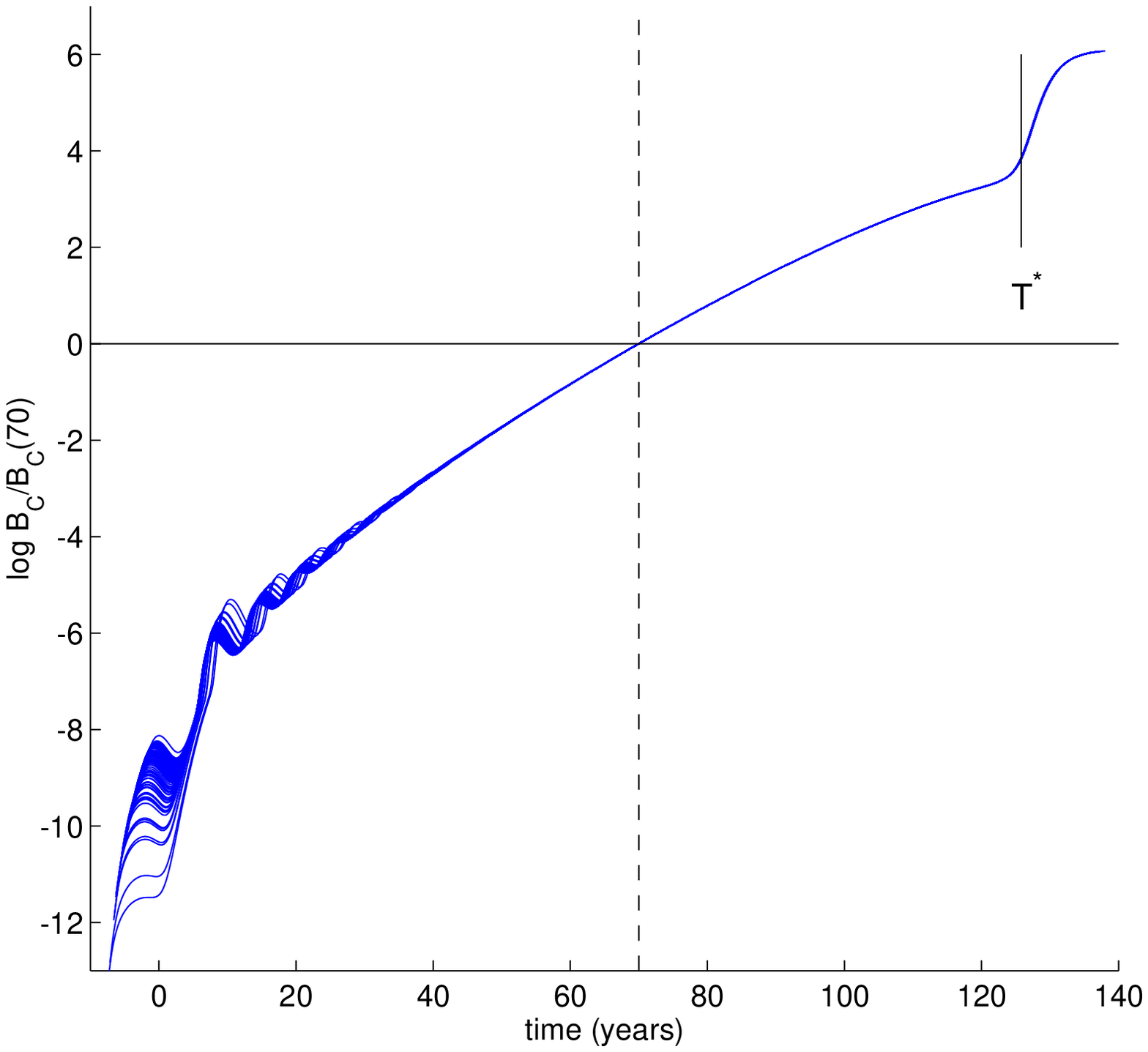}}

\begin{caption}{Sensitivity of the solution to perturbation of the 
initial conditions.}
\label{fig:timeshift}
\end{caption} 
\end{center} 
\end{figure}

Figure~\ref{fig:timeshift} is instructive in this regard. Here one hundred
simulations are plotted with the eight initial conditions in
Table~\ref{table:IC} for each run perturbed by a fractional change of
$\delta_k$ for $k=1\ldots 8$, where $\delta_k$ is a normal random variable
of zero mean and standard deviation $\sigma = 0.01$.  The results are
plotted each with an individually adjusted time shift $ \Delta t =
c_2/\Im(\mu_0)$ where $c_2$ is the phase in (\ref{eq:bca}),
computed by a log least squares fit on the range $t =[60,90]$.  This amounts to
synchronizing the time in each realization by using for a reference the
asymptotic exponential solution to define the origin, $t=0$.  With this
timing shift, the average of all simulations is equivalent to
starting at $t = -3.98$ years and with a standard deviation of $\sigma =
1.12$ years.  It takes, in other words, on average about four years to get
locked into a long period cycle of growth for initial conditions in this
neighborhood.

For clarity, each realization of $B_C$ is also scaled by the amplitude of
the associated {\sl asymptotic} solution (\ref{eq:bca}) at $t=70$ in the
shifted frame, that is, $ | B_C^{(0)}| \, \exp(\Re(\mu_0)\, (70-\Delta t))$.  That
scaling is significant because the ensemble of computed values of $B_C$
spans two orders of magnitude by $t=120$, a huge sensitivity to initial
conditions indeed! But now the asymptotic rescaling and shift make it
abundantly clear that the onset of collapse is rigidly locked in time on
adopting (\ref{eq:bca}) to define the time origin of the growth cycle.
The latter synchronization is a manifestation of nonlinearity and can be
seen as a consequence of the associated matched asymptotic expansion; the
collapse cannot be adjoined to the growth phase at an arbitrary point in
the cycle. Rather, the required smoothness of connections simultaneously
in not just $B_C$, but all eight variables, singles out a unique point,
late in the cycle, where the match must be made. The piece that connects
the two exponential solutions contains a subdominant contribution that
reflects the influence of particular initial conditions. The conclusion is
that the leading order sensitivity of the solution timing to initial
conditions is solely the varying interval it takes for locking onto the
growth cycle. Once this is achieved, the timing of the collapse is
preordained.

On this view one should describe collapse as taking one of two asymptotic
mathematical forms; the immediate collapse and the deferred collapse. The
first is the solution given in (\ref{eq:crstart}-\ref{eq:crend}). The
second is a more involved structure, a so-called ``uniform asymptotic
expansion,'' consisting of the solution stated in
(\ref{eq:bceq}-\ref{eq:lameq}) from \S \ref{sect:unstable}, using canonical
initial conditions at $t=0$ (scale amplitudes and phases for the
particular parameters, as in Table~\ref{table:ivp}, with free parameters of
$[ B_C^{(0)}, \Delta t_{B_C}, K_r^{(0)}]$ determined by the eight initial
conditions),\footnote{Note that phase shifts as those in
  Table~\ref{table:ivp} are not absolute, but all {\sl relative} to
  $\Delta t_{B_C}$. But $K_r$ does {\sl not} have any shift at all.}  and
switching at late time to (\ref{eq:crstart}-\ref{eq:crend}) by means of a
correction term that effects the match to leading order.\footnote{The
  earlier depiction of that correction term as a simple exponential
  arising from an eigenvalue computation is too crude an approach for the
  match indicated; one has rather to discover a certain structure, a
  so-called internal boundary layer, on the graphical evidence a problem
  suited to the WKBJ method for a first-order turning point.} It is then
only a question of which of these two solutions is the ``nearer" for a
given set of initial conditions in the eight-dimensional 
space of Table 2.\footnote{From this perspective, 
one can argue the real exponential solution has
to be finite amplitude stable simply because no matched asymptotic expansion to the collapsed state is possible.}

With regard to duration of the economic cycle as a function of variation
in the initial conditions, this is a ``regular'' perturbation; the value
of $\sigma = 1.12$ is about a one percent change in the realized value of
$T$, of the same order as the scale of the perturbation of initial
conditions. Of course the definition of $T$ is somewhat ambiguous
depending what characteristic feature of the solution at late time is
chosen but an average value of $130$ years is plausible and this differs
modestly from the leading order estimate of $\pi/\Im(\mu_0)$.

\begin{figure}
\begin{center} 
\epsfxsize=3.5in{\epsfbox{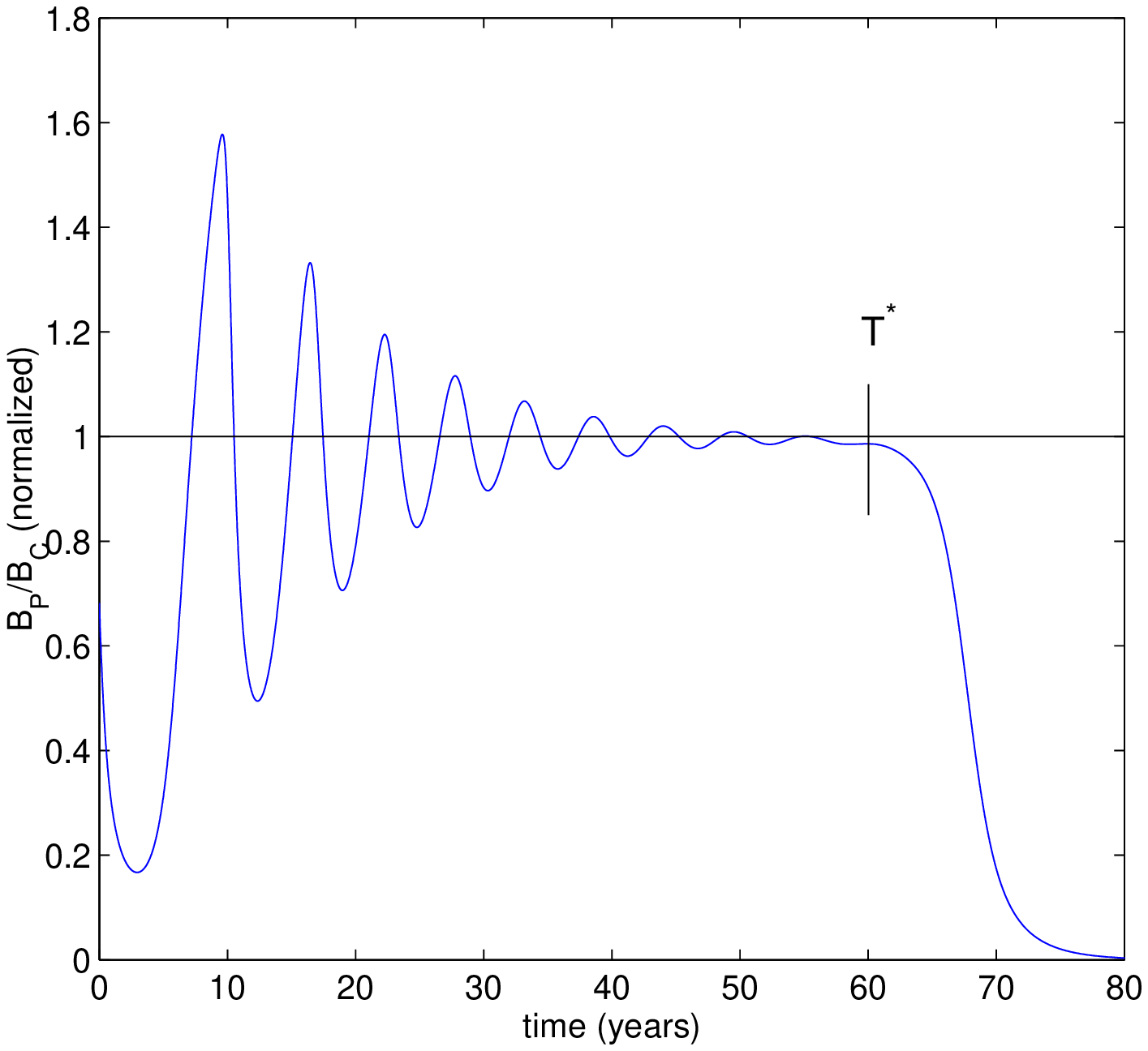}}
 
\begin{caption}{A shorter period of growth, with collapse still accurately
indicated by $T^*$.}
\label{fig:short_case}
\end{caption} 
\end{center} 
\end{figure}

Note from the plot that the phase-adjusted refinement of this, $T^*$, is
an excellent predictor for the transition to collapse. This accuracy holds
up for shorter cycles as well, as shown in Figure~\ref{fig:short_case},
with parameters varied from Table~\ref{table:standard} in a Monte Carlo
run.  This parameter set results in 60 years of growth before collapse,
similar in duration to the postwar private sector debt run-up from 1945
through 2008. The plot is similar to Figure~\ref{fig:flfd}, with another
time-lagged ratio, $B_P/B_C$, normalized so that the predicted
asymptotic limit for the growth phase is again unity.

\subsection{Regime diagrams}
\label{sect:regime}

\begin{figure}
\begin{center} 
\epsfxsize=3.5in{\epsfbox{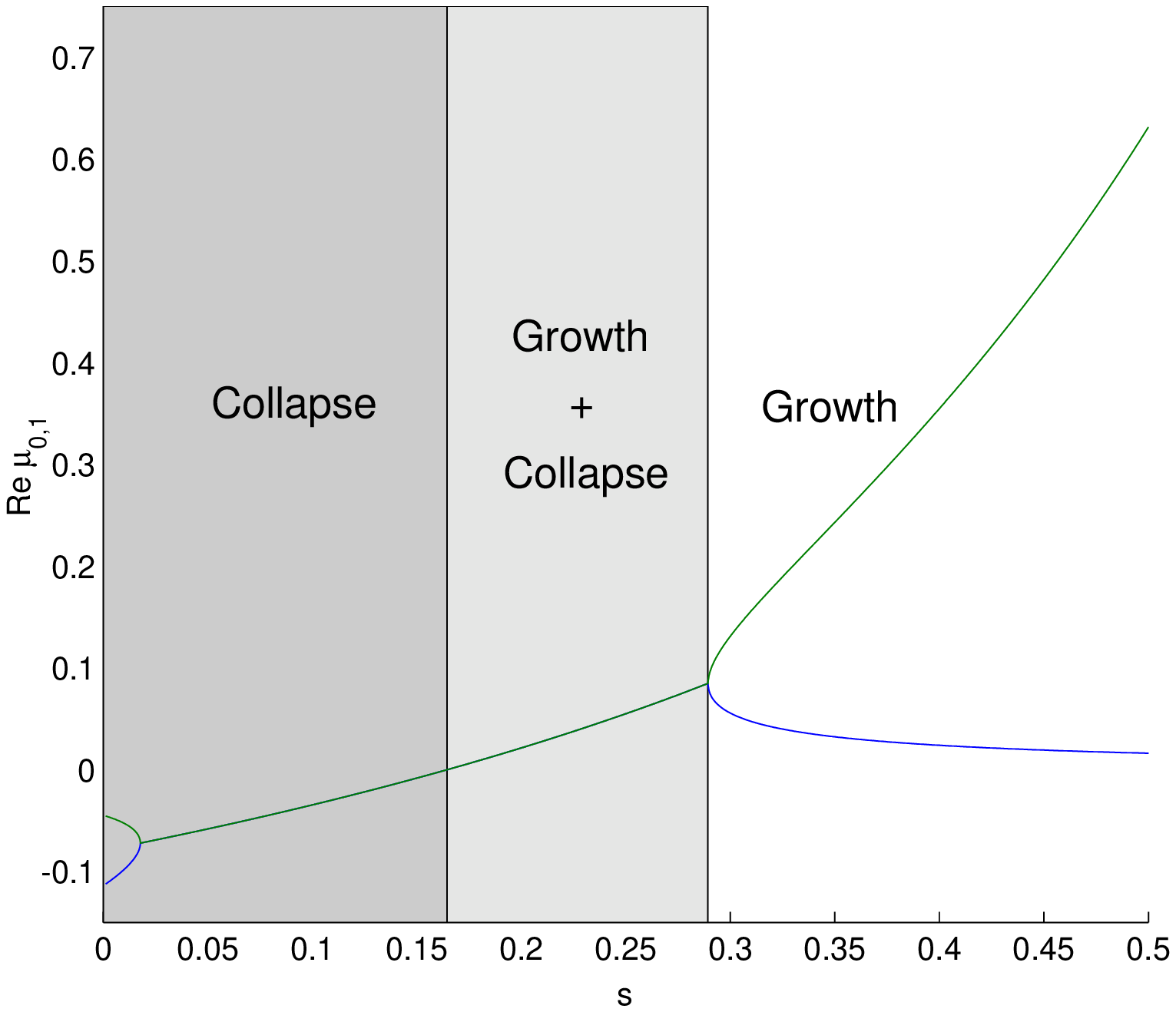}}

\begin{caption}{Bifurcation from stability to instability induced by the
return to capital, $s$.}
\label{fig:regime}
\end{caption} 
\end{center} 
\end{figure}

Returning once more to the focus on the single bifurcation parameter $s$,
we can by aid of (\ref{eq:sbif}) quickly sketch out a regime diagram,
shown in Figure~\ref{fig:regime}. To the right, the dominant root of $\mu$
is real and the solution of \S \ref{sect:stable} is the outcome for most
initial conditions. In the middle (light gray), the leading root is a
complex pair and the usual outcome is the ``deferred collapse'', with a
period of growth (\S \ref{sect:unstable}) followed by a quick collapse
(\S \ref{sect:collapse}). To the left (dark gray), all roots of $\mu$ lie
in the left half plane so no exponential growth is possible and all
initial conditions lead to collapse. (Note that the apparent bifurcation for
small $s$ is nugatory. The two leading roots are real however the largest
of these is still less than zero, violating the assumption that the exponential
terms dominate, e.g.\ the constant sum of $F_L - F_D - B_{PL} - W_D$.)

One would like to understand the general sensitivity of this system
to bifurcations such as this one in $s$ (or $v$). The order of the
characteristic polynomial for $\mu$, quintic, is a function of the
form of the differential equations; modifying parameters cannot change
that. But as with $s$, so too variations of other parameters may induce
bifurcations. There are a large number here and in the remainder we
briefly examine the influence of ten: $[\tau_B, \tau_W, r_L, \tau_{PC},
  r_D, s, v, \alpha, \delta, \beta]$.  (It would be possibly instructive
--- and feasible --- to enrich this space with sixteen more; the
parameters in the four generalized exponential functions for $\tau_{RL}$,
$\tau_{LC}$, $\mbox{Inv}(\pi_r)$, and $\mbox{Ph}(\lambda)$.)

This is rather too much freedom to exploit and it is useful first to
undertake a more limited exploration; to vary each of the ten parameters
singly and compare results. A natural way to make this comparison is to
replace a candidate variable $ x$ whose value in the standard model is
$x_0$ with $\nu \, x_0 $, and allow $\nu$ to vary from $1/4$ to $4$ in
equal logarithmic steps. For a number of these, a factor of four either
way lies well outside the range of real world plausibility but the
exploration is for the moment a purely formal one.

\begin{figure}
\begin{center} 
\epsfxsize=4in{\epsfbox{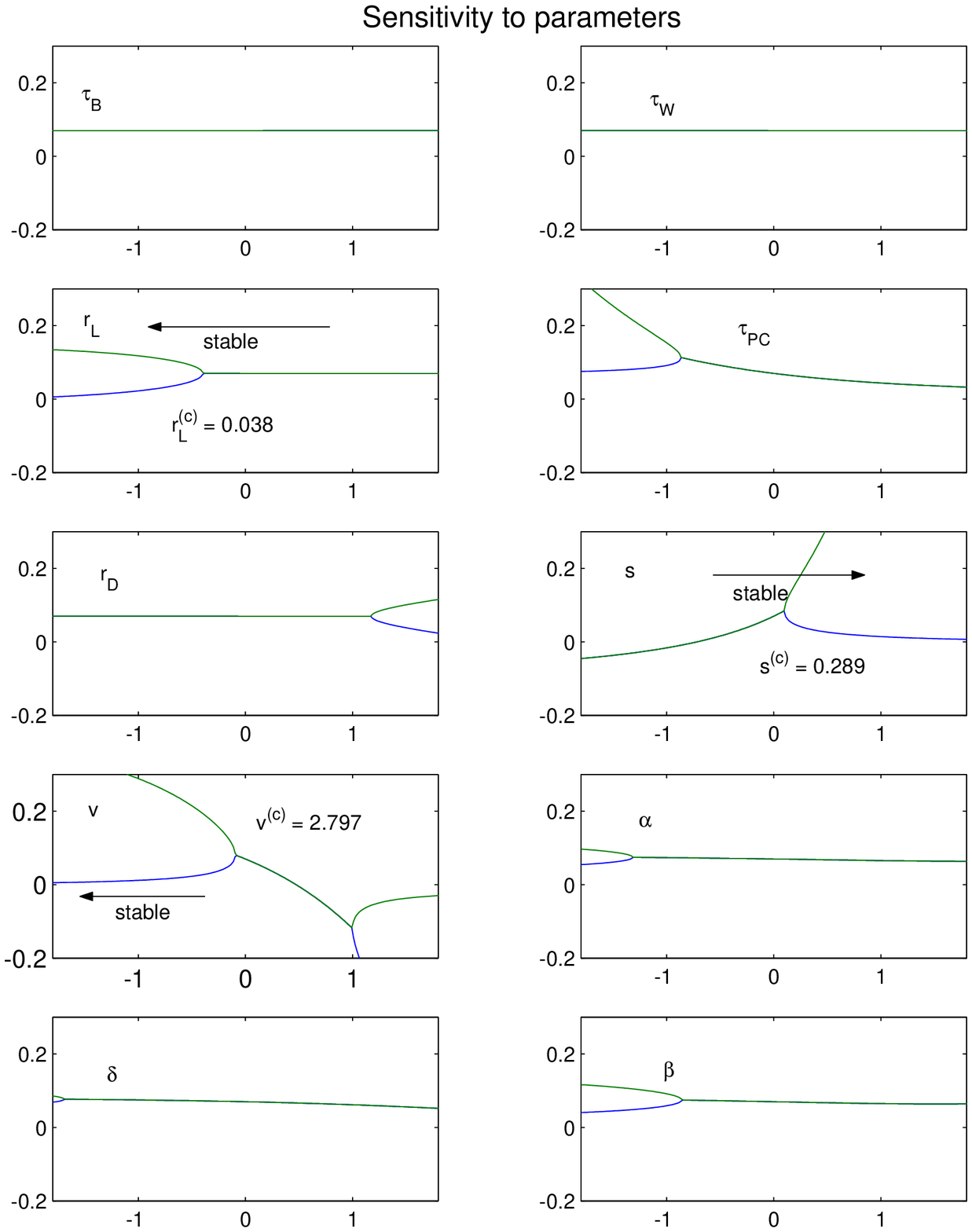}}

\begin{caption}{Sensitivity of bifurcations to the ten model parameters individually}
\label{fig:sensitive}
\end{caption} 
\end{center} 
\end{figure}

The results in Figure~\ref{fig:sensitive} show that the parameters
$\tau_B$ and $\tau_W$ for example have negligible influence on the
roots. The three that are most sensitive are the return to capital $s$,
with a bifurcation to stability for {\sl increasing} $s$ at $0.289$, for a
{\sl decreasing} capital-output ratio $v$ at $2.797$, and a {\sl
  decreasing} interest rate on loans $r_L$ at $0.038$. (Each of these
values is readily verified with short time stepping runs a bit beyond the
point of bifurcation, confirming that the ten parameter coding of the
characteristic polynomial is correct.)  Other parameters, such as
$\tau_{PC}$ do induce a bifurcation, but only for a relatively large
change in value (about a factor of two). 

\begin{figure}
\begin{center} 
\epsfxsize=4in{\epsfbox{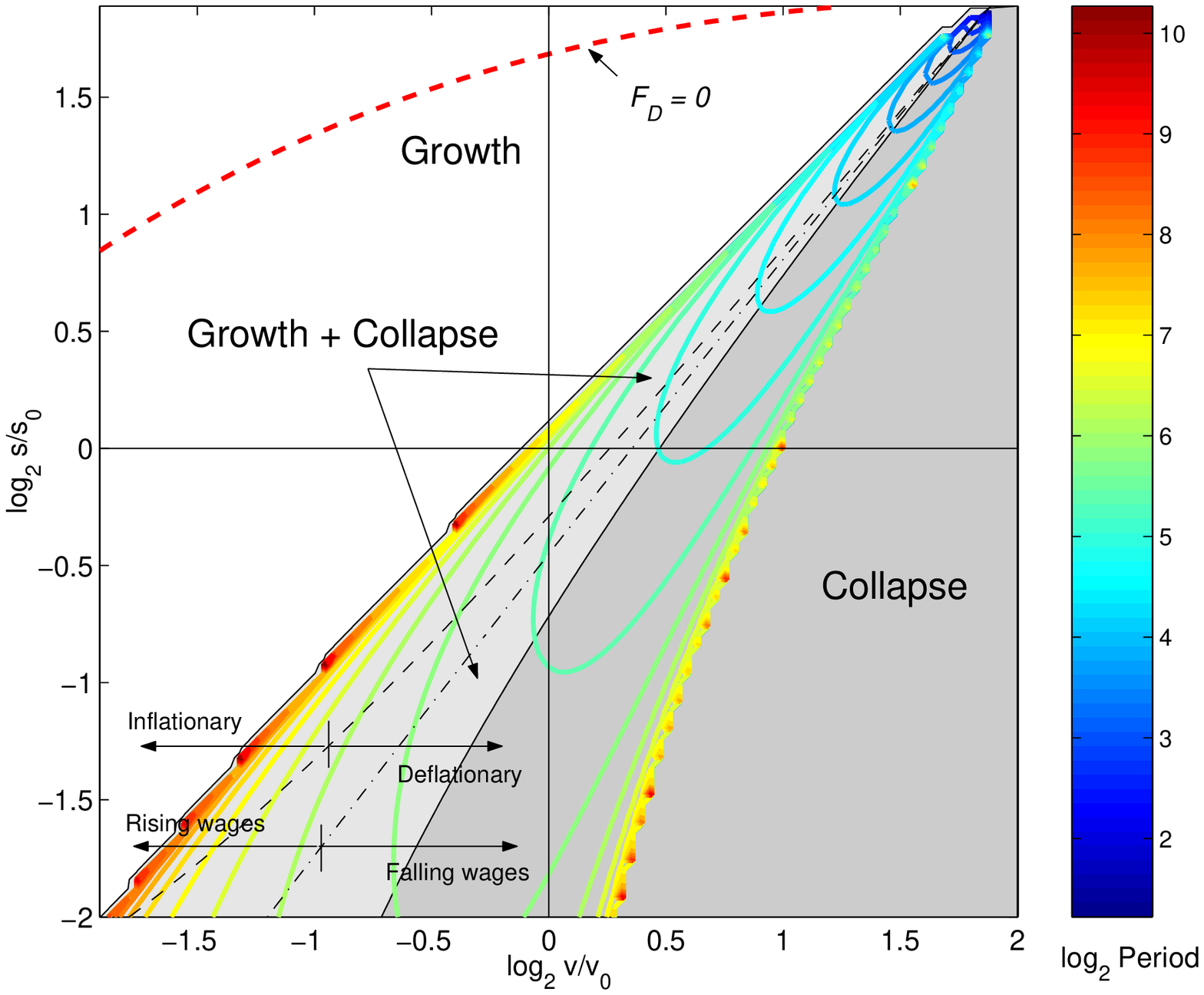}}

\begin{caption}{Bifurcation in the two-dimensional subspace of 
$[s,v]$, with shading as in Figure~\ref{fig:regime}.
Contours indicate the approximate
period, $T$, of  economic growth but these apply only
in the light gray region.}
\label{fig:planar}
\end{caption} 
\end{center} 
\end{figure}

While $r_L$ is a relevant factor, for both simplicity and graphical reasons
we confine attention
to a two-dimensional subspace, that of $[s, v]$. Figure~\ref{fig:planar}
illustrates the border of stability in this plane, with axes as those
in Figure~\ref{fig:sensitive}. The vertical line corresponds to
the slice shown in panel six of Figure~\ref{fig:sensitive}, the horizontal
line to panel seven. Their intersection is the standard model. The log
of the leading order estimate of $T = \pi/\Im(\mu_0)$ is contoured in color.
Note that the odd behavior in the upper right hand corner, where the period
becomes extremely short, reflects the return to capital $s$ approaching
unity. 

The border separating stable from unstable growth makes intuitive
sense. Its positive slope shows that, as the economy becomes more capital
intensive (increasing $v$), the minimal return to capital, $s_{crit}$,
that ensures stability is monotone increasing. Near the center of the
plot, that relation is approximately $s_{crit} = 1.12\, s_0\,
(v/v_0)^{1.07}$.  Moreover, there is a critical value of $v$ above which
the economy must always collapse directly.

Setting aside phase lags in the complex case, when $\Re(\mu) = \beta +
\alpha$, from (\ref{eq:price}) prices are asymptotically constant. This is
shown by the dashed line. To the left the economy is inflationary, prices
of goods rise; to the right deflationary.  So in this model, and quite
likely much more generally for debt-based economies, inflation is
stabilizing. Furthermore workers here enjoy a rising standard of living
(until a collapse), with wages in constant dollars growing in lockstep
with productivity gains, as in the US from 1947 to 1973. This holds even
in a deflationary regime and with falling wages (to the right of the
dash-dotted line), albeit the period of growth then becomes very short.

Also note in this figure the red dashed line that demarcates a critical
state where $F_D = 0$, the firm has no assets at all (but the bank
continues to advance it loans on the Micawberish promise of future
earnings). This is an intrinsic model limit; values of $(s,v)$ above the
red line lead to negative asymptotic values for $F_D$.  As remarked
earlier, realizable models in general are those for which the entire set
$[B_{PL}^{(0)}, F_L^{(0)}, F_D^{(0)}, W^{(0)}, P_C^{(0)} ]$ yields
non-negative results.  In the particular slice shown here it simply
happens to be $F_D^{(0)}$ that defines the border of allowed parameter
space.

Below this line in the allowed region, firm assets ($F_D$) are
always less than firm liabilities ($F_L$) --- the firm perpetually runs in
the red.  This is not unexpected in the (complex) regime of mixed growth
and collapse where, as in Figure \ref{fig:flfd}, the appropriate phase lag
is used, but it remains true even in the pure growth regime, where there
is no phase lag. But recall that at leading order the linear
coefficients satisfy $ W_D^{(0)} +  F_D^{(0)} =  F_L^{(0)} - B_{PL}^{(0)}$ and
since all terms are constrained to be positive definite for a consistent
model, then necessarily $ F_D^{(0)} \le F_L^{(0)}$. If worker loans were 
introduced to the model we would have only the weaker aggregate inequality
that $  W_D^{(0)} +  F_D^{(0)} \le F_L^{(0)} + W_L^{(0)}$ and one or the other 
could have assets greater than liabilities, but not both.
 
Reincorporating $r_L$, the wedge-shaped region typified by deferred
collapse becomes a volume whose varying extent in that third dimension
could next be explored. With the addition of further parameters, a
complicated hypervolume results but, again, the main operative constraints
on stability as suggested by Figure~\ref{fig:sensitive} remain
the subspace spanned by $[s,v,r_L]$.

For completeness sake it should be noted that the border separating light
from dark gray, $\Re(\mu_0) = 0$, constitutes another class of solutions,
these exhibiting algebraic, rather than exponential, behavior.  Also in
this spirit are solutions that lie at borders in the space of initial
conditions as noted in \S \ref{sect:stable}, rather than in parameter
space.  But border solutions of either kind are of limited relevance 
as they are unstable to an infinitesimal perturbation.


\subsection{Model Robustness}
\label{sect:robust}

Another aspect of the model bears comment. One would like to be sure that
the identification of stable and unstable regimes is a robust one. Here we
can look to the forms chosen for $a(t)$ and $N(t)$ from concern that the
arguments given earlier, which trade heavily on the properties of
exponential functions, point to an implausible sensitivity on having
exactly an exponential form for the underlying variables $a$ and $N$. While
one can argue that an exponential growth model is justified for population
(excepting inconveniences such as wars and plagues), there is not really
much to be said in its defense for $a(t)$ except that historical data show
a concave upward trend.

So one might instead try a 
replacement for both of these in the form of algebraic growth:
\[
  a  = a_0\, (t_0+\alpha\, t)^{\nu}\qquad
  N  = N_0 \, (t_0+\beta\, t)^{\nu}\, .
\]
By suitable adjustment of the exponent and constants, 
each of these can be made to give
curves similar enough to data  over a time span of say, $50$ years,
that a model based on these latter forms had better give qualitatively
similar results. (The choice of $\nu = 5/2$ is reasonable for matching
the standard case.)

Briefly, this form does largely recapitulate the results above (and with
surprisingly little change in e.g., $s_{crit}$). One can conjecture that
the weak requirement of merely a concave upward form for $a$ and $N$
suffices. One could strengthen this argument for robustness further by
looking to equivalent conclusions for variation in the functional form of
$G$, the all purpose exponential here for $\tau_{RL}$ and allied
functions, but there are not likely to be any surprises in that direction
either.

\section{Conclusions}
\label{sect:conclusion}

At times mathematical manipulations can obscure, not reveal, the meaning
of a model. So it bears emphasizing that the central points here are
simply expressed. Any economy has a finite capacity for carrying aggregate
debt. How much debt depends upon various factors, notably among them the
three earlier cited: interest rate, return to capital, and the capital to
output ratio. In a stable Schumpeter economy all the relevant economic
indicators march together in synchrony. When the debt burden {\sl exceeds}
the carrying capacity of the economy, the hallmark is the necessary lag of
key variables. We have seen this in Figure \ref{fig:flfd}. Recall that
deposits lag loans and only by reason of that delay is it possible to
sustain speculative investment in that model economy.

It is evident this relation must be unidirectional in time; it would be
nonsensical were the time ordering reversed. So much is clear with no need
of mathematical argument. Indeed, judging from the byzantine result of
Appendix A for the phase of wages, $W$, it is likely not even possible to
prove such a phase relation between $F_L$ and $F_D$. And for the model
here it may not even be true for all possible parameter values. But, were
loans to lag deposits, we should simply conclude the model was in that
parameter range irrelevant to the real world.\footnote{Falling back on
  numerical results, this lag was determined for each member of a Monte
  Carlo simulation consisting of 33,000 models clustered about the
  parameter values in Table \ref{table:standard}, allowing independent
  normal (Gaussian) fractional variations of zero mean and a standard
  deviation of ten percent. The shortest lag observed was about three
  weeks, the longest 5 months. While this is by no means a proof, it at
  least confirms one's intuition about positivity.}
 
So Schumpeter's ``creative destruction" is a healthy aspect of capitalism
only when the economy in which it operates has adequate capacity to
sustain the associated debt and that does not automatically come about
from the invisible hand of Adam Smith. It requires no great leap of faith
to conclude that Ponzi finance must inevitably overwhelm the carrying
capacity of {\sl any} economy.

To be sure, the model here is at best a ``lumped parameter'' description
of a complicated nonlinear system. One is mindful from the Sonnenschein
Mantel Debreu theorem that macroeconomic modeling has in key respects only
a tenuous relation to microeconomic models and so it is fair to question
how robust are the conclusions drawn here; whether the equations that
compose this model are even an appropriate representation of aggregate
behavior. The best line of argument in defense is probably to be had from
appeal to empirical time series.

Michael Hudson has written often of the central role of debt in economies
since antiquity, particularly underscoring Marx's incisive critique of
debt in capitalism, emphasizing its exponential growth \cite{Hudson_exp}
as well as its ratcheting quality: ``... every U.S.\ business recovery
since WW II has taken off with a higher level of debt. That means debt
service --- the monthly `nut' you have to cover --- has become much higher
with each recovery.'' \cite{Hudson_ratchet} The ultimately catalytic role
of mounting debt in precipitating collapse is pungently expressed by
Hudson in his epigram ``debts that can't be paid won't
be.''\cite{Hudson_debt}

Keen has expressed this dynamic through the time rate of change of debt
contributing to the aggregate demand. And what emerges then in the system
here defined by (\ref{eq:sys1} -- \ref{eq:sys8}) is rapid evolution onto a
manifold, to use the nomenclature of nonlinear dynamical systems.

The typical occurrence of ``slow'' manifolds as illustrated in textbook
problems for a system of two or three first-order nonlinear differential
equations has one identify variables as either ``fast'' or ``slow''. The
former are said to be ``slaved'' to the slow variables, and this signifies
that equations for the fast variables have the time derivative eliminated,
leaving an algebraic relation that defines fast in terms of slow. This
relation fixes in turn the structure of the slow manifold, normally by an
iterative process that may or may not give a convergent global
representation.  Commonly the fast variables evolve exponentially in time
while evolution {\sl on} the manifold is algebraic, so the disparity in
time scales is great.

Here the fast evolution is exponential in time, but so is the slow
evolution; there is only a difference in rates and hence the decaying
epicycles as in Figure~\ref{fig:short_case} remain apparent.  The approach
here aims also at a exposing a manifold, but the resulting uniform
asymptotic expansion of \S \ref{sect:collapse} is more correctly thought
of as a closely related ``inertial'' manifold.

That aim has largely been achieved. The one technical omission, of a
detailed match of two exponential solutions, is not likely to be very
revealing. It would give a more refined value for $T^*$, but one hardly
differing from the conceptually much simpler, and substantially accurate,
phase-corrected value already defined.

What emerges in a broader sense is an understanding that the manifold
reflects the centrality of debt in economic systems.  The economy is
indeed ``slaved" to debt, not merely in the narrow technical sense of
dynamical systems, but in human terms as well. Atop this structure lie the
epicycles that offer the great economic sport of daily life, the engaged
players blithely unaware of a slower clock, ticking away inexorably on a
multi-decadal time scale, the hidden alarm long ago set.

\bibliographystyle{elsarticle-harv}
\bibliography{econref}	

\appendix

\section{ Correction for $T^*$}
\label{sect:appendixA}

Typically the largest phase shift, which dictates the corrected period $T^*$,
is that for $W$. In terms of the complex root $\mu_0$, that
phase shift can be computed exactly according to:
\[
W^{(0)} =  \frac{\mu_0 \, (s-1)}{\alpha + \beta + \delta} \, \left (1 + \tau_{P_C}\, \left (\mu_0 -\alpha - \beta\right )\right ) \, \frac{N(\mu_0, \gamma)}{D(\gamma)}
\]
where
\begin{eqnarray*}
N(\mu_0, \gamma) &=&  6125 \, (5\, {\rm e}^2)^{1/7} \,(\mu_0 + 2)\, \gamma^{26}  +
175\, (5^4\, {\rm e})^{1/7}\, (6\, \mu_0 + 13)\, \gamma^{20}
+ 75\, (3\, \mu_0+ 7) \, \gamma^{14} \\
\qquad && + 735\, \mu_0 \, (5^{17}/{\rm e})^{1/21}\, \gamma^{12}
+ 105\, (5^5 / {\rm e}^4)^{1/21}\, (6\, \mu_0 + 1)\, \gamma^6
 + 9 \, (5^2/{\rm e})^{1/3}\, (3\, \mu_0 + 1)\, ,\\
D(\gamma) &=& (7 \, (5^4\, {\rm e})^{1/7} \, \gamma^6 + 3)\, 
(25 \, \gamma^{14} +  3\, (5^2 / {\rm e})^{1/3}) \, ,\\
\gamma &=& \left (v\, \left (\alpha + \beta + \delta\right )\right )^{1/21}\, .
\end{eqnarray*}
The phase shift is then $\tan^{-1} ( \Im(W^{(0)})/\Re(W^{(0)}))$ and
the scale amplitude is $|W^{(0)}|$.

\section{Full asymptotic expansion; Origin of the Great Moderation}
\label{sect:appendixB}

A more complete version of (\ref{eq:bceq} -  \ref{eq:lameq}) assumes the form:
\begin{eqnarray}
\tilde B_C  &\equiv&  B_C  - \frac{\tau_{LC}^{(0)}}{\tau_{RL}^{(0)}} \, \frac{r_D\, cst}{r_D-r_L} \sim
 B_C^{(0)}\, \left ( \exp(\mu_0\, t)+B_C^{(1)}\, \exp(\nu_1\, t)
+B_C^{(2)}\, \exp(\nu_2\, t) 
\right. \nonumber\\
&&\quad\quad \left. 
+B_C^{(3)}\, \exp( (2 \nu_1 - \mu_0) \, t) 
 +B_C^{(4)}\, \exp( (\nu_1 + \nu_2 - \mu_0) \, t) +
\ldots \right ) \label{eq:asymp_begin}\\
  B_{PL}&=& B_C^{(0)}\, \left( B_{PL}^{(0)}\, \exp(\mu_0\, t)+B_{PL}^{(1)}\, 
\exp(\nu_1\, t) +B_{PL}^{(2)}\, \exp(\nu_2\, t) 
 \right. \nonumber\\
&&\quad\quad \left. 
+B_{PL}^{(3)}\, \exp( (2 \nu_1 - \mu_0) \, t)
+B_{PL}^{(4)}\, \exp( (\nu_1 + \nu_2 - \mu_0) \, t) + \ldots \right )\\
\tilde F_L &\equiv&  F_L - \frac{r_D\, cst}{r_D-r_L} \sim
 B_C^{(0)}\, \left ( F_L^{(0)}\, \exp(\mu_0\, t)+F_L^{(1)}\, \exp(\nu_1\, t)
+F_L^{(2)}\, \exp(\nu_2\, t) 
                     \right. \nonumber\\
&&\quad\quad \left. 
+F_L^{(3)}\, \exp( (2 \nu_1 - \mu_0) \, t) 
+F_L^{(4)}\, \exp( (\nu_1 + \nu_2 - \mu_0) \, t) + \ldots \right )
\label{eq:asymp_fl}\\
\tilde F_D &\equiv& F_D -  \frac{r_L\, cst}{r_D-r_L} \sim
 B_C^{(0)}\, \left ( F_D^{(0)}\, \exp(\mu_0\, t)+F_D^{(1)}\, \exp(\nu_1\, t)
+F_D^{(2)}\, \exp(\nu_2\, t)    \right. \nonumber\\
&&\quad\quad \left. 
+F_D^{(3)}\, \exp( (2 \nu_1 - \mu_0) \, t) 
+F_D^{(4)}\, \exp( (\nu_1 + \nu_2 - \mu_0) \, t) + \ldots \right ) \label{eq:asymp_fd}\\
  W   &=& B_C^{(0)}/K_r^{(0)}\, \left ( W^{(0)}\, \exp((\mu_0 - \beta)\, t)
                   +W^{(1)}\, \exp((\nu_1-\beta)\, t)  +W^{(2)}\, \exp((\nu_2-\beta)\, t) \right. \nonumber \\
      && \qquad \qquad \left.
+W^{(3)}\, \exp((2 \nu_1 - \mu_0 -\beta)\, t) 
+W^{(4)}\, \exp((\nu_1 + \nu_2 - \mu_0 -\beta)\, t) + \ldots \right )\\
  P_C &=& B_C^{(0)}/K_r^{(0)}\, \left ( P_C^{(0)}\, \exp((\mu_0 - \beta - \alpha)\, t)
                  +P_C^{(1)}\, \exp((\nu_1 - \beta - \alpha)\, t)
+P_C^{(2)}\, \exp((\nu_2 - \beta - \alpha)\, t)  \right. \nonumber\\
 && \left. 
+P_C^{(3)}\, \exp((2 \nu_1 - \mu_0 - \beta - \alpha)\, t) 
+P_C^{(4)}\, \exp((\nu_1 + \nu_2 - \mu_0 - \beta - \alpha)\, t) 
+ \ldots \right )\\
 K_r &=& K_r^{(0)} \, \exp((\alpha+\beta)\, t)\, \Bigl (1
+   \frac{9\,\pi_r^{(1)}}{4\, (\nu_1-\mu_0)} \,  \exp((\nu_1-\mu_0)\, t)
+\frac{9\,\pi_r^{(2)}}{4\, (\nu_2-\mu_0)} \,  \exp((\nu_2-\mu_0)\, t) 
 \nonumber \\
&&  
+\frac{9\,\pi_r^{(3)}}{4\, (2 \nu_1-\mu_0)} \,  \exp((2 \nu_1-\mu_0)\, t) 
+\frac{9\,\pi_r^{(4)}}{4\, (\nu_1 + \nu_2-\mu_0)} \,  \exp((\nu_1 + \nu_2-\mu_0)\, t) + \ldots
\Bigr )\\
\lambda &=& \lambda^{(0)}\, \Bigl ( 1
+\frac{9 \,  \pi_r^{(1)}}{4\, (\nu_1-\mu_0)} \, \exp((\nu_1-\mu_0)\, t)  
+\frac{9 \,  \pi_r^{(2)}}{4\, (\nu_2-\mu_0)} \, \exp((\nu_2-\mu_0)\, t) \nonumber\\
&&
+\frac{9 \,  \pi_r^{(3)}}{4\, (2 \nu_1-\mu_0)} \, \exp((2 \nu_1-\mu_0)\, t)
+\frac{9 \,  \pi_r^{(4)}}{4\, (\nu_1 + \nu_2-\mu_0)} \, \exp((\nu_1 + \nu_2-\mu_0)\, t) 
+ \ldots \Bigr)\, .\label{eq:asymp_end}
\end{eqnarray}
The significant generalization here is that quantities previously assumed
constant now incorporate the transient corrections as well, as seen in the
last expression above for $\lambda(t)$. The defining transient is taken to
be in the variable $\pi_r$, now written in the form
\be
\pi_r   =  \pi_r^{(0)}  +   \pi_r^{(1)}\, \exp((\nu_1-\mu_0)\, t) 
+   \pi_r^{(2)}\, \exp((\nu_2-\mu_0)\, t) 
+   \pi_r^{(3)}\, \exp((2 \nu_1-\mu_0)\, t) +  
 \pi_r^{(4)}\, \exp((\nu_1 + \nu_2-\mu_0)\, t) + \ldots \, .
\ee
From this latter, the corresponding transient amplitudes in both $\lambda$
and $K_r$ above are immediately expressed in terms of $\pi_r^{(k)}$ as
indicated. Similar transient forms are easily deduced for the auxiliary
quantities $[g, \tau_{RL}, \tau_{LC}, Inv, Ph]$. Though the variables
$\pi_r^{(k)}$ are ultimately eliminated in favor of $B_C^{(k)}$ by use of
(\ref{eq:profit}), which defines $\pi_r$, they are extremely useful as an
intermediate placeholder for a tractable formulation of the problem.

Here for simplicity we restrict attention to the case of real 
$\mu_0$. (The analysis is similar for the complex case.)  Each of
(\ref{eq:asymp_begin}-\ref{eq:asymp_end}) is a formal asymptotic infinite
series about $t \to \infty$. While the radius of convergence of an
asymptotic series is zero, such series are nonetheless extremely useful,
just not all the way to the origin $t=0$ in the present case.

The earlier noted degeneracy of equations for $d W/dt$ and $d P_C/dt$ is
true only in leading order. At next order each must separately be
enforced.  The requirement that the series above hence satisfy
(\ref{eq:sys1} -- \ref{eq:sys8}) and (\ref{eq:profit}) at second order yields
an eleventh degree polynomial in $\nu$ whose coefficients depend, not only
on the explicit parameters in Table 1, but on the value of $\mu_0$ as
well. Four of the roots of this polynomial are always $\mu_0$ but
these are discarded, so the reduced polynomial is seventh degree.

This polynomial is too complicated to carry along the general ten
parameter dependence of the coefficients, even with the aid of symbolic
manipulation, hence we turn now to the particular case of $s=0.3$
considered in \S \ref{sect:stable} for which the leading roots are found
to be
\[
\nu_1 = 0.05496705 \qquad 
\nu_2 = 0.04054906 + 1.22315328 \, i \, .
\]
Of the seven roots of $\nu$, four are simply slight perturbations of the
quintic roots noted in the main body of the paper.  Here for comparison
e.g., $\mu_1 = 0.05540080$. But $\nu_2$ and its conjugate companion have
no match from leading order; these are new spontaneous solutions of the
system. The next two roots are $\nu_4 = -0.99$ (exactly) and $\nu_5 =
-0.99082495$. The first of this pair can be matched with $\mu_3$. The
second lacks a match however one observes that perturbing a constant
solution and allowing for a single common exponent correction for all
variables yields a double root of $-1$, that is, solutions of $\exp(-t)$
and $t \, \exp(-t)$, so the pair $\nu_{4,5}$ would seem to be a splitting
of this repeated root. Finally $\nu_{6,7}$ match with $\mu_{4,5}$.

Where the leading order asymptotic expansion in \S \ref{sect:anal} had
only two free real parameters, $[ B_C^{(0)}, K_r^{(0)} ]$, the form above
has the added $ B_C^{(k)} $ that accompany the single exponent form
$\exp(\nu \, t)$ (note that $B_C^{(2)}$, is complex, hence equivalent to
two real parameters; amplitude and phase).  This development of the
polynomial in $\nu$ seems to result in one too many free parameters; seven
from the new roots and the two scale parameters introduced in \S
\ref{sect:anal}.  The resolution is that dynamics of the equation for
$W_D$ have been implicitly incorporated through the accounting identity
noted at the beginning of the paper and so nine free parameters are
required for a general solution.  The expansion above has been truncated
at $\nu_2$ both for relative compactness of the formulae and because the
remaining terms in $\nu_k$ are negligible but formally they must be
included to constitute a general solution that {\sl is} an asymptotic
expansion of an exact solution of (\ref{eq:sys1} -- \ref{eq:sys8}).  This
exponent partition reflects that the {\sl effective} dimension of the
system is not nine, but five.\footnote{The initial two parameters of \S
  \ref{sect:anal} and accompanying value of $\mu_0$ play roughly the role
  of what, in the parlance of asymptotic expansions, is termed the
  ``controlling factor'', while the five parameter truncation is the more
  proper analog of the ``leading order'' behavior.}  For the truncated
form above, coefficients of remaining terms with superscripts $(1,2)$ and
of {\sl all} terms with superscripts $(3,4)$ can be found in terms of the
five parameters $[ B_C^{(0,1,2)}, K_r^{(0)} ]$.

Note that (\ref{eq:asymp_begin}), (\ref{eq:asymp_fl}), and
(\ref{eq:asymp_fd}) address the initial assumption of
homogeneity, which led to the identification of the scale parameters $[
  B_C^{(0)}, K_r^{(0)} ]$.  Properly incorporated, the conserved quantity
redefines the relevant variables for asymptotic expansion, i.e., the
equations have to be recast in terms of $[ \tilde B_C, \tilde F_L,
  \tilde F_D ]$.

There is a final subtlety to be noted about the expansion.  If one takes
the explicit ninth order system defined by (\ref{eq:sys1} - \ref{eq:sys8})
plus the equation for $dW_D/dt$ and follows the same analysis, a different
polynomial emerges. Terming those roots $\rho_k$, then $\rho_{1,2,3} =
\nu_{1,2,3}$ but there is an added root $\rho_4 = 0$, which reflects that
there is a conserved quantity for the equations, here absorbed in the
redefined tilde variables. In addition $\rho_{7,8} = \nu_{5,6}$ but
remaining roots $\rho_{5,6}$ (a repeated real root) and $\nu_{4,7}$
differ. Both pairs are at the level of rapidly decaying transients, not in
the effective five-dimensional subspace. The discrepancy is resolved by
additional terms stemming from recasting the equations in terms of $[
  \tilde B_C, \tilde F_L, \tilde F_D ]$ to reflect the conservation law, 
leading to a modification of $\nu_{4,7}$.




Because the general asymptotic solution cannot be carried back to the
origin, one could not directly relate a given set of nine initial
conditions to an equivalent set of values for the free parameters. This is
a minor cavil. One could always evolve the solution from $t=0$ to some
moderate value $t_0$, where the truncated asymptotic series were judged
sufficiently accurate and make the connection there. More to the point is
that the mapping from the values of the variables onto the (now) nine free
parameters, whether at $t_0$ or at the origin, is a nonlinear one and, as
already signaled by Figure~\ref{fig:bistab}, there is no such real-valued
mapping for a certain continuous region in the space of nine variable
values. That region constitutes the space of initial conditions that map
instead to the collapse solution.

\begin{figure}
\begin{center} 
\epsfxsize=4in{\epsfbox{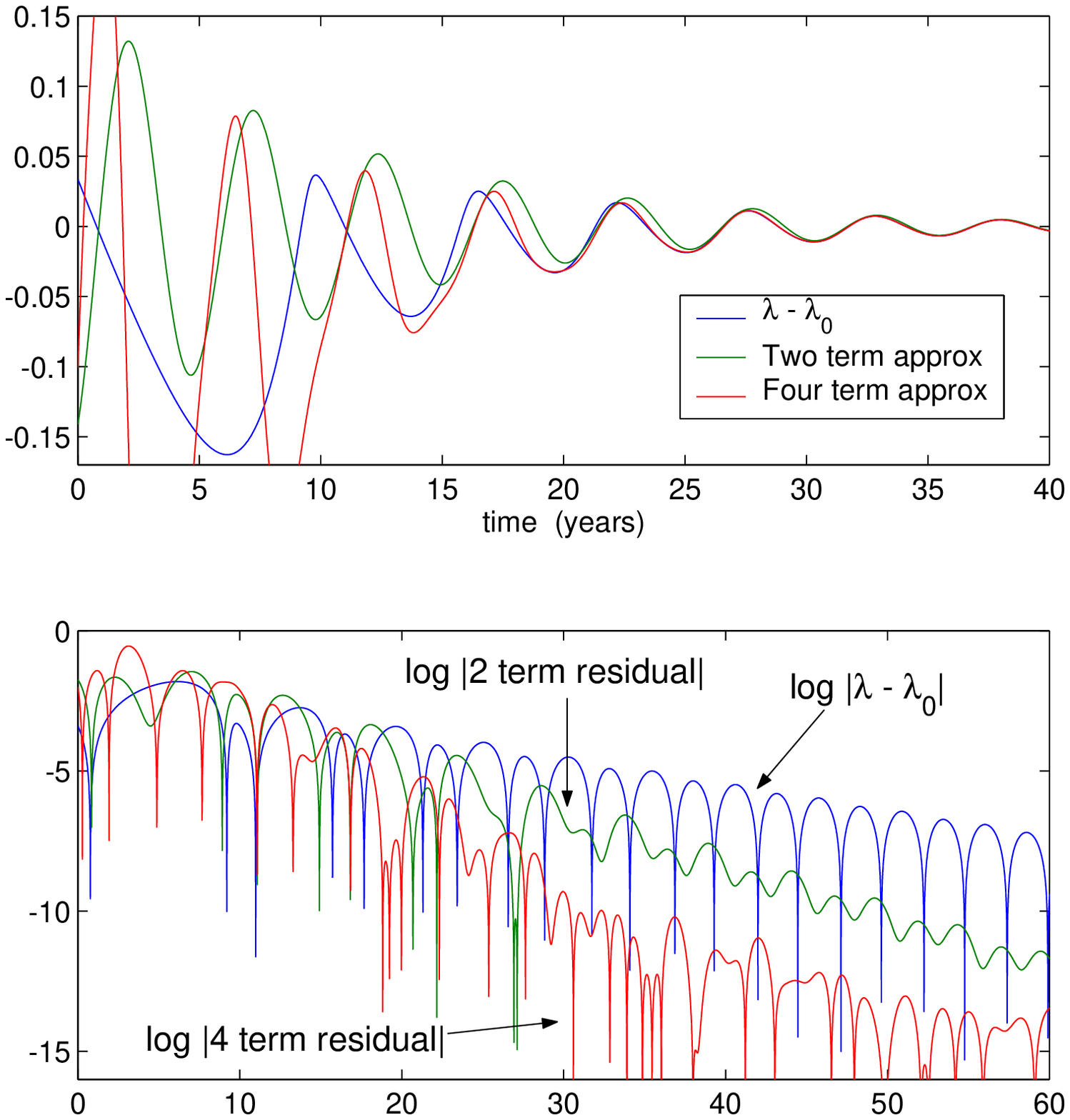}}

\begin{caption}{Higher order corrections in the asymptotic expansion}
\label{fig:moderation}
\end{caption} 
\end{center} 
\end{figure}

To most easily see the first few correction terms emerge it is convenient to test
(\ref{eq:asymp_end}). As usual, this requires an empirical determination of
constants, here $\pi_r^{(1,2)}$ (with $\pi_r^{(3,4)}$ found by a
recurrence relation).  The first two assume the values
\[
\pi_r^{(1)}= 4.0481601 \times 10^{-5}, \quad 
| \pi_r^{(2)}| = 6.7677858 \times 10^{-3}, \quad 
\phi = 2.0915926\, 
\]
where $\phi$ is the phase shift, as in previous instances of the complex
exponential. The accuracy of (\ref{eq:asymp_end}) is evident in
Figure~\ref{fig:moderation}. From the top panel one would say that the
solution has locked onto the asymptotic solution at this level by fifteen
years. Comparison of two- and four-term expansions suggests that, if a few
more terms are kept, reasonable agreement might be pushed back to the ten
year mark.\footnote{Strictly by ordering of exponents, the cross-term $\exp((\nu_1 +
  \nu_2 - \mu_0) t) $ precedes $\exp((2 \, \nu_2 - \mu_0) t)$, but the
  latter happens to be numerically more significant for any $t$ within
  reason and this is the fourth term used for the plots. In any event with
  asymptotic expansions there is always an optimal number to
  retain. Thereafter the results get worse, though this can sometimes be
  surmounted by resummation.  Here the form $B_C^{(0)}\, \exp ( \mu_0\, t +
  b_C^{(1)}\, \exp((\nu_1-\mu_0) \, t) + b_C^{(2)}\, \exp((\nu_2 - \mu_0)
  \, t) )$ shows promise.} The log of the two-term residual, plotted in
the lower panel clearly shows the frequency doubling of $2\, \nu_2$ and
this is largely captured with the four-term result.

For both stable and unstable growth cycles there is always a spontaneous
complex root pair $\nu_{2,3}$, as above, with a period typically of the
order of a few years.  This is the origin of the Great Moderation
exhibited by the Keen Model.  While by more abstract argument such a
qualitative phenomenon may more easily be deduced, insofar as one wishes a
precise determination of the period and growth rate, these values
evidently cannot emerge from any more elementary analysis than that given
here, which rests upon a particular balance of higher order terms.  As
$\Re(\nu_2)> 0$, this indicates a growing transient. But, as incorporated
above, there is either a real offset of $\nu_2 - \mu_0$ in the same
exponential (as for $\lambda$) or else a separate term growing faster as
$\mu_0$, against which background this growth loses ground, hence the
effect is always one of relative decay.

It is the rare nonlinear ninth-order system that admits a general
solution and, by means of that, access to a parametric
understanding of all essential aspects of the system.  It seems this is a
peculiarity of debt-driven economic systems and the present elementary
approach to determine a hierarchy of exponents should thus be borne in
mind when considering other models of this sort.

\end{document}